\pdfoutput=1
\documentclass{ar-1col}
\usepackage[comma]{natbib}
\usepackage{url}
\usepackage{amsmath}
\usepackage{amssymb}
\usepackage{multirow}
\jname{Xxxx. Xxx. Xxx. Xxx.}
\jvol{AA}
\jyear{YYYY}
\doi{10.1146/((please add article doi))}

\begin{document}

\markboth{In\'acio et al.}{Statistical Evaluation of Medical Tests}

\title{Statistical Evaluation of Medical Tests}

\author{Vanda In\'acio,$^1$ Mar\'ia Xos\'e Rodr\'iguez-\'Alvarez,$^2$ and Pilar Gayoso-Diz$^3$
\affil{$^1$ School of Mathematics, University of Edinburgh, EH9 3FD Edinburgh, United Kingdom; email: vanda.inacio@ed.ac.uk}
\affil{$^2$BCAM-Basque Center for Applied Mathematics \& IKERBASQUE, Basque Foundation for Science, E-48009 Bilbao, Spain}
\affil{$^3$Instituto de Salud Carlos III, 28029 Madrid, Spain}}

\begin{abstract}
In this review, we present an overview of the main aspects related to the statistical evaluation of medical tests for diagnosis and prognosis. Measures of diagnostic performance for binary tests, such as sensitivity, specificity, and predictive values, are introduced, and extensions to the case of continuous-outcome tests are detailed. Special focus is placed on the receiver operating characteristic (ROC) curve and its estimation, with the topic of covariate adjustment receiving a great deal of attention. The extension to the case of time-dependent ROC curves for evaluating prognostic accuracy is also touched upon. We apply several of the approaches described to a dataset derived from a study aimed to evaluate the ability of HOMA-IR (homeostasis model assessment of insulin resistance) levels to identify individuals at high cardio-metabolic risk and how such discriminatory ability might be influenced by age and gender. We also outline software available for the implementation of the methods.
\end{abstract}

\maketitle


\section{INTRODUCTION}
Evaluating and ranking the performance of medical tests for screening and diagnosing disease greatly contributes to the health promotion of individuals and communities. Throughout this article we will be using the term `diagnostic test' to broadly include any continuous classifier, such as, a single biological marker or a univariate composite score obtained from a combination of biomarkers. The primary goal of a diagnostic test is to distinguish between individuals with and without a well-defined condition (termed `disease', with `nondisease' used to indicate the absence of the condition). For some diseases, there might exist a gold standard test that perfectly classifies all individuals as diseased or nondiseased. However, gold standard tests (e.g., a biopsy) might not only be expensive, but also invasive and potentially harmful.  Economic and/or ethical reasons may thus preclude the routine use of gold standard tests except when sufficient evidence is present. As a consequence, much effort has been placed in developing new candidate tests that are less invasive, costly, or easier to apply than the gold standard counterpart. Nevertheless, new candidate tests are rarely perfect. Thus, a critical step prior to approving the use of a diagnostic test in clinical practice is to rigorously vet its ability to distinguish diseased from nondiseased individuals.  Compared to the truth, i.e., to the diagnosis made by the gold standard test, which we assume to be available, interest lies in quantifying the misclassification errors made by the test under investigation and in deciding whether yet with such errors, the test may still be suitable for routine use. It is worth noting that although we focus on medical diagnosis, the problem of binary classification is such a wide one, finding applications in fields as diverse as finance (e.g., customer likely to incur in default or not) and cyber security (e.g., email messages are spam or not), to name only two.  

The receiver operating characteristic (ROC) curve \citep{Metz78} is the most popular used tool for evaluating the discriminatory ability of continuous-outcome tests, which are our focus. ROC curves thus receive a great deal of attention in this article. The ROC curve was developed during World War II to assess the ability of radar operators to differentiate signal (e.g., enemy aircraft) from noise (e.g., flock of birds). Its expansion to other fields was prompt (e.g., psychology) and it was first extensively used in radiology to evaluate medical imaging devices \citep{Metz86}. Thanks to advancements in technology, with a vast array of ways to diagnose disease or to predict its progression available and with new diagnostic tests or biomarkers continuously being studied, the ROC curve is, nowadays, a key tool in medicine. ROC curves are also widely used in machine learning to evaluate classification algorithms.  Quoting \citet[][p.~1]{Gneiting18} there has been an `\emph{(...) astonishing rise in the use of ROC curves in the scientific literature. In 2017, nearly 8,000 papers were published that use ROC curves, up from less than 50 per year through 1990 and less than 1,000 papers annually through 2002.}'.

The aim of this article is to present an overview of the main statistical concepts and methods for evaluating the accuracy of medical tests, with ROC curves naturally receiving the main emphasis. The reader is referred to the books by \cite{Pepe03}, \cite{Krzanowski09}, \cite{Zhou11}, \cite{Broemeling16} and papers cited in this article for further coverage of the topic. 

The remainder of this article is structured as follows: In Section \ref{ilu} we describe the HOMA-IR dataset, which is used as an illustrative example throughout the article. Measures of diagnostic accuracy, including the ROC curve and some methods for its estimation, are introduced in Section \ref{acc_measures}. The topic of covariate-adjustment in ROC curves is reviewed in Section \ref{covariateroc}, while in Section \ref{timeROC} time-dependent ROC curves are discussed. In Section \ref{software} we outline available software in \texttt{R} \citep{R20}. Finally, in Section \ref{discussion}, we offer some conclusions and thoughts on further topics.
\section{ILLUSTRATIVE EXAMPLE}\label{ilu}
Insulin resistance (IR) is a feature of disorders such as type 2 diabetes mellitus and is implicated in obesity, hypertension, cancer, or autoimmune diseases. Also, IR is associated with cardiovascular diseases, and some studies have shown that IR may be an important predictor of cardiovascular disease risk. The HOmeostasis Model Assessment of IR (HOMA-IR) is widely used in epidemiological studies and in clinical practice to estimate IR and has proved to be a robust tool for the surrogate assessment of IR. We will exemplify some of the different measures and methods described in this paper when it comes to studying the capacity of HOMA-IR levels to detect patients with higher cardio-metabolic risk and to ascertaining the possible effect of both age and gender on the accuracy of this measure. The purpose here is merely illustrative, and we refer the interested reader to \cite{Gayoso13}, where the objective was originally proposed and studied, for more details and references.

In particular, as an accurate indicator of the presence of cardio-metabolic risk (i.e., presence of `disease'), we use a diagnosis of metabolic syndrome as defined by the International Diabetes Federation \citep{IDF20} criteria, under which metabolic syndrome is defined as the presence of central obesity (defined as waist circumference with ethnicity specific values) plus any two of the following four risk factors: (1) reduced HDL-cholesterol or specific treatment for this lipid abnormality, (2) raised systolic or diastolic blood pressure or treatment of previously diagnosed hypertension, (3) raised fasting plasma glucose or previously diagnosed type 2 diabetes, (4) raised triglycerides or specific treatment for this lipid abnormality.

\begin{figure}[h]
\includegraphics[width=13cm]{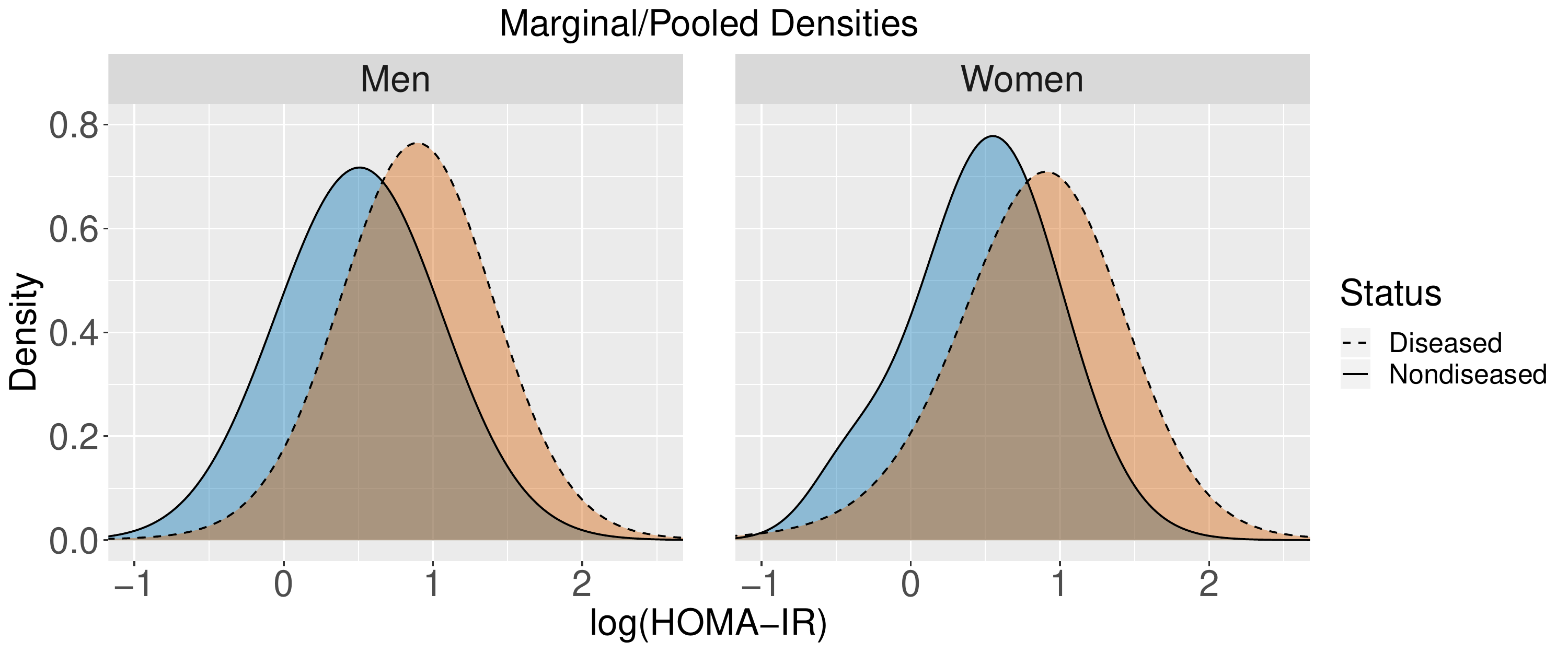}
\caption{Estimated density functions of $\log$ HOMA-IR levels obtained by fitting a Dirichlet process mixture of normals model to each population and separately for men and women.}
\label{estdensities}
\end{figure}

Regarding the study population, it corresponds to the individuals enrolled in the EPIRCE study (Estudio Epidemiol\'ogico de la Insuficiencia Renal en Espa\~na) \citep{Otero05, Otero10}, which is an observational cross-sectional study that included a randomly selected sample of Spanish individuals aged $20$ years and older, stratified by age, gender, and residence. For the analyses shown here, $2212$ individuals out of $2459$ were selected (age range in years $20$--$92$). Subjects with diabetes ($247$, $10.0\%$ of the total sample) were excluded. Of the total of $2212$ subjects, $41.0\%$ were men ($769$ nondiseased and $135$ diseased) and $59.0\%$ women ($1194$ nondiseased and $114$ diseased). All participants were Caucasians. Table~\ref{EPIRCEDataDescriptive} presents some summary statistics of the HOMA-IR levels (log-transformed) for men and women, as well as, for different age strata. In turn, Figure~\ref{estdensities} depicts, separately for men and women, the estimated density functions of the $\log$ HOMA-IR levels in the diseased and nondiseased populations. As can be observed, both in men and women, individuals with metabolic syndrome tend to have higher HOMA-IR levels and these levels also vary with age.
  
\begin{table}
\tabcolsep7.5pt
\caption{Median (interquartile range) of the (log) HOMA-IR levels in diseased and nondiseased populations, males and females, and for four gender strata based on quartiles.}
\centering
\begin{tabular}{lcc}
& \textbf{Diseased} & \textbf{Nondiseased}\\\hline
\textbf{Global sample} & $0.91\;(0.50, 1.25)$ & $0.51\;(0.13, 0.85)$\\
\hline
\textbf{Gender} & &\\
Women & $0.89\;(0.49, 1.26)$ & $0.51\;(0.14, 0.82)$\\
Men & $0.92\;(0.52, 1.25)$ & $0.50\;(0.11, 0.89)$\\\hline
\textbf{Age} & &\\
$\leq 35$ & $1.06\;(0.64, 1.34)$ & $0.53\;(0.18, 0.85)$\\
$(35, 47]$ & $1.04\;(0.69, 1.35)$ & $0.47\;(0.09, 0.82)$\\
$(47, 60]$ & $0.87\;(0.52, 1.19)$ & $0.47\;(0.08, 0.81)$\\
$> 60$  & $0.82\;(0.42, 1.25)$ & $0.60\;(0.17, 0.92)$\\
\hline
\end{tabular}
\label{EPIRCEDataDescriptive}
\end{table}
\section{POPULAR MEASURES OF ACCURACY}\label{acc_measures}
\subsection{Binary Tests}
Although our focus is on tests measured on a continuous scale, we start by defining measures of classification accuracy for binary tests as they provide the natural starting point for what comes next. A binary test is a test for which  there are only two possible outcomes, usually denoted as positive or negative for the condition or disease of interest. Let $Y$ be a binary variable denoting the diagnostic test outcome, with $Y=1$ indicating a positive test result for disease, and $Y=0$ indicating a negative test result for disease. Further, let $D$ be the binary variable that denotes the true disease status, and let $D=1$ denote the presence of disease and $D=0$ indicate its absence. The accuracy of a test is defined as its ability to distinguish between diseased and nondiseased individuals and can be measured by its true positive and true negative fractions. The true positive fraction of a test, TPF, also known as sensitivity, is the probability that a diseased individual tests positive, that is, $\text{TPF}=\Pr(Y=1\mid D=1)$. The true negative fraction, TNF, also known as specificity, is the probability that a nondiseased subject tests negative, i.e., $\text{TNF}=\Pr(Y=0\mid D=0)$. The ideal test would correctly classify all diseased and nondiseased individuals, but the tests routinely used in practice are relatively inexpensive and classification errors do occur. Specifically, two types of misclassification are possible: a diseased individual can test negative and a nondiseased individual can test positive. The magnitude of such misclassification errors is measured through the false negative fraction (FNF) and the false positive fraction (FPF), which are defined as, $\text{FNF}=\Pr(Y=0\mid D=1)$ and $\text{FPF}=\Pr(Y=1\mid D=0)$. Clearly, $\text{FNF}=1-\text{TPF}$ and $\text{FPF}=1-\text{TNF}$. An ideal test is one for which the TPF and TNF are both equal to one or, equivalently, where the FNF and FPF are both equal to zero. Obviously, the closer such quantities are to these ideal values, the more the classification made by the test is to be trusted. Nevertheless, a test can be useful even when these quantities are smaller than the ideal values. The criterion whereby the validity of a test is established in practice depends entirely on the context in which it is to be applied. For example, a false negative outcome can be life-threatening with diseased individuals failing to receive prompt treatment while, on the other hand, a false positive outcome may result in the physical, emotional, and financial burdens resulting from further testing or even unnecessary treatment.

The true positive and negative fractions quantify how well the test performs among subjects with and without the condition, respectively, which is important for public health concerns. In the clinical setting, however, interest resides in the opposite question, i.e., how well the test outcome predicts the true disease status. The question of interest is: Given that an individual has a positive (negative) test outcome, what is the probability of being diseased (nondiseased)? This leads to the positive and negative predictive values (PPV and NPV, respectively)
\begin{align}
\text{PPV}&=\Pr(D=1\mid Y =1)=\frac{\pi\text{TPF}}{\pi\text{TPF}+(1-\pi)\text{FPF}}, \label{ppv}\\
\text{NPV}&=\Pr(D=0\mid Y =0)=\frac{(1-\pi)\text{TNF}}{(1-\pi)\text{TNF}+\pi\text{FNF}}, \label{npv}
\end{align}
where $\pi=\Pr(D=1)$ is the prevalence of the disease in the source population. An ideal test has PPV and NPV both equal to 1, that is, it predicts disease status perfectly. On the other hand, for a noninformative test one has that $\text{PPV}=\pi$ and $\text{NPV}=1-\pi$, i.e., the test has no information about the true disease status or, in other words, information about the test outcome is independent of disease status. Since the predictive values depend on the prevalence of the disease, their interpretation must be cautious. For instance, a low PPV may be due to a low disease prevalence or to a test that poorly reflects the true disease status.

It has been suggested (e.g., \citealt[][Chapter 2]{Pepe03}) to use the TPF and FNF for quantifying the inherent accuracy of a test, as these classification probabilities quantify how well a given test reflects true disease status. Predictive values, in turn, quantify the clinical or practical value of the test, rather than its accuracy. That is, diagnostic accuracy must refer to the quality of the information yielded by the test (i.e., its TPF and TNF), something that has to be distinguished from the usefulness or practical utility of such information (quantified by the predictive values). It is worth mentioning at this stage that as the TPF and TNF are independent of disease's prevalence, they can be estimated from case-control studies. By opposition, estimation of the predictive values requires that the prevalence is known or that it can be estimated from the data.
\subsection{Continuous Tests}\label{continuous}
\begin{figure}[h]
\includegraphics[width=13cm]{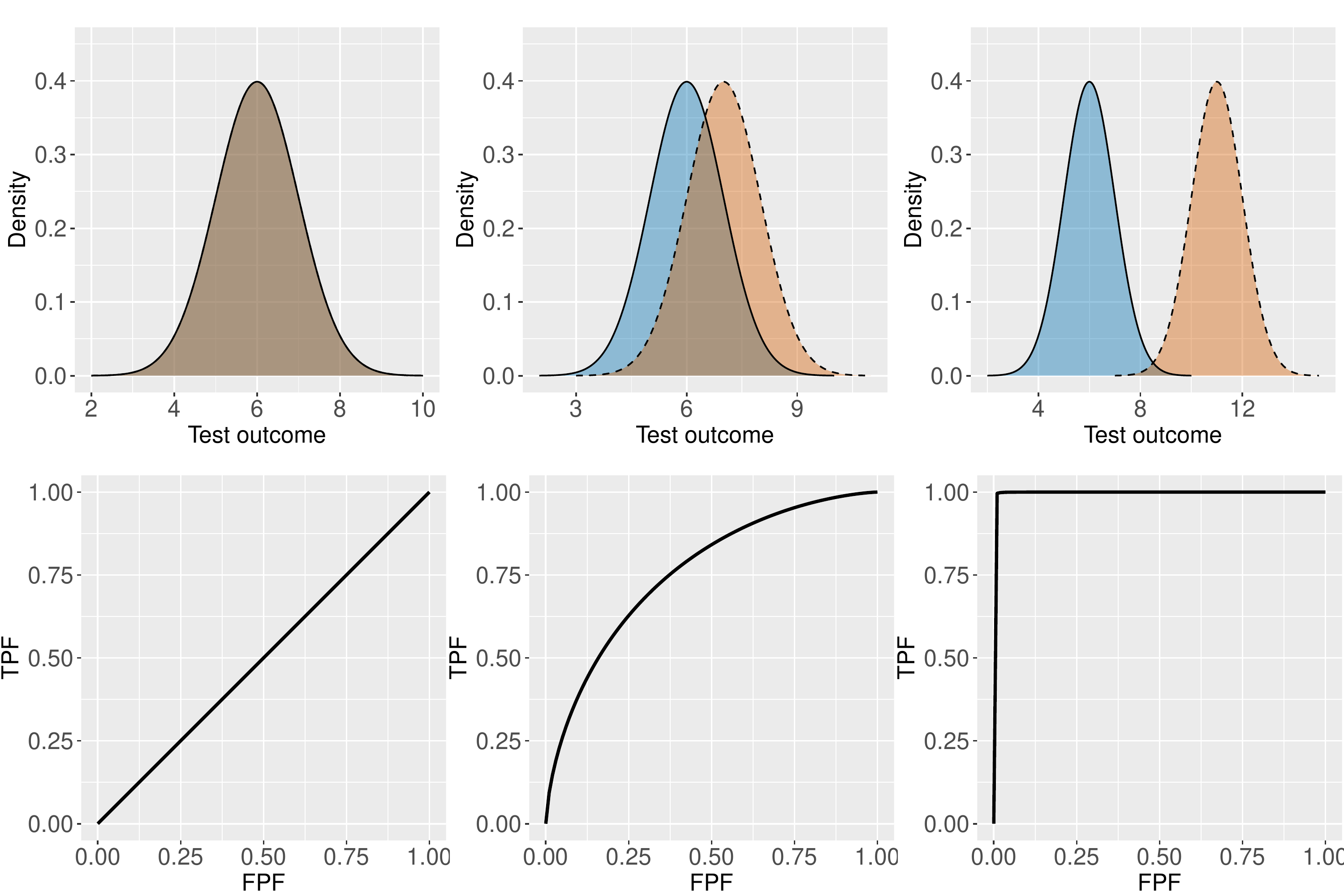}
\caption{Hypothetical densities of test outcomes in the diseased (dotted line, orange) and nondiseased (solid line, blue) populations (top) along with the corresponding ROC curves (bottom).}
\label{densrocs}
\end{figure}

Although some tests are naturally dichotomous, such as commercial home pregnancy tests or bacterial cultures, many tests are continuous (e.g., HOMA-IR levels for predicting the presence of cardio-metabolic risk). The question arising is how to classify an individual as diseased or nondiseased based on his/her test result, which is now measured on a continuous scale. The simplest classification is based on a cutoff or threshold value, say $c$, such that a test result with $Y\geq c$ is considered positive for disease and if $Y<c$ the test is considered negative. Therefore, each threshold value chosen gives rise to a corresponding TPF and TNF, or equivalently, to a TPF and FPF, that is,
\begin{align*}
\text{TPF}(c)&=\Pr(Y\geq c\mid D=1)=\Pr(Y_D\geq c)=1-F_D(c),\\
\text{FPF}(c)&=\Pr(Y\geq c\mid D=0)=\Pr(Y_{\bar{D}}\geq c)=1-F_{\bar{D}}(c),
\end{align*}
where we use the subscripts $D$ and $\bar{D}$ to index related quantities to the diseased ($D=1$) and nondiseased ($D=0$) populations, and with $F_D$ and $F_{\bar{D}}$ denoting the cumulative distribution function of test results in the diseased and nondiseased populations, respectively. It is clear that there will be as many pairs of true and false positive fractions as of threshold values chosen and comparing all of them would be impractical. This leads us to the popular ROC curve, which represents nothing more than the plot of the FPF versus the TPF as the threshold value used for defining a positive test result is varied, that is
\begin{equation*}
\{(\text{FPF}(c),\text{TPF}(c)):c\in\mathbb{R}\}=\{(1-F_{\bar{D}}(c),1-F_{D}(c)):c\in\mathbb{R}\}.
\end{equation*}
The ROC curve thus provides a visual description of the tradeoff between the FPF and TPF as the threshold $c$ changes. For $p=\text{FPF}(c)=1-F_{\bar{D}}(c)$, the ROC curve can be equivalently represented as
\begin{equation}\label{rocdef}
\{(p,\text{ROC}(p)):p\in [0,1]\},\quad \text{with}\quad \text{ROC}(p)=1-F_{D}\{F_{\bar{D}}^{-1}(1-p)\}.
\end{equation}
Further advantages afforded by the ROC curve as a measure of a test's accuracy are that: (a) it is not dependent on disease prevalence, (b) it is independent of the units in which diagnostic test results are measured, thereby enabling ROC curves of different diagnostic tests, and thus their diagnostic accuracy, to be compared, and (c) it is invariant to strictly increasing transformations of the diagnostic test result $Y$. We shed some light on how ROC curves should be interpreted. ROC curves measure the amount of separation between the distribution of test outcomes in the diseased and nondiseased populations (see Figure \ref{densrocs}). When the distributions of test results in the two populations completely overlap, then the ROC curve is the diagonal line of the unit square, with $\text{FPF}(c)=\text{TPF}(c)$ for all $c$, indicating a noninformative test. The more separated the distributions of test outcomes, the closer the ROC curve is to the point $(0,1)$ and, consequently, the better the diagnostic accuracy. A curve that reaches the point $(0,1)$ has $\text{FPF}(c)=0$ and $\text{TPF}(c)=1$ for some threshold $c$ and, hence, corresponds to a test that perfectly determines the true disease status. An ROC curve which lies below the diagonal line implies that the test is worse than useless, but this issue can be easily overcome by reversing the classification rule, i.e., to say that an individual is diseased when his/her test outcome is below $c$ and nondiseased otherwise. Related to the ROC curve is the notion of placement value \citep{Pepe04}, which is simply a standardisation of test outcomes with respect to a reference population. Let $U_D=1-F_{\bar{D}}(Y_D)$ be the placement value of diseased individuals with respect to the nondiseased population. This variable $U_D$ quantifies the degree of separation between the diseased and nondiseased populations. Specifically, if test outcomes in the two populations are highly separated, the placement of most diseased individuals is at the  upper tail of the nondiseased distribution and so most of them will have small $U_D$ values. In turn, if the two populations overlap substantially, $U_D$ will have a $\text{Uniform}(0,1)$ distribution. Interestingly, the ROC curve turns out to be the cumulative distribution function of $U_D$, that is, $\Pr(U_D\leq p)=\text{ROC}(p)$.

A standard way to summarise the information provided by the ROC curve is to calculate the area under the ROC curve (AUC), which is defined as
\begin{equation*}
\text{AUC}=\int_{0}^{1}\text{ROC}(p)\text{d}p.
\end{equation*}
In addition to its geometric definition, the AUC has also a probabilistic interpretation \citep[see, e.g.,][p. 78]{Pepe03}
\begin{equation}
\text{AUC} = \Pr\left(Y_{D} \geq Y_{\bar{D}}\right),
\label{AUC2}
\end{equation}
that is, the AUC is the probability that the test outcome for a randomly chosen diseased subject exceeds the one exhibited by a randomly selected nondiseased individual. The AUC is equal to 1 for a perfect test and it is equal to $0.5$ for a test with no discriminatory power (see Figure~\ref{densrocs}). Another global summary measure of diagnostic accuracy is the Youden index(YI) \citep{Youden50}, defined as
\begin{align}
\text{YI}&=\max_c\{\text{TPF}(c)+\text{TNF}(c) -1\} \nonumber \\
&=\max_c\{F_{\bar{D}}(c)-F_D(c)\}\label{YIcdf} \\
&=\max_p\{\text{ROC}(p)-p\}\label{YIroc}.
\end{align}
When the distributions of test outcomes completely overlap $\text{YI}=0$, whereas when they are completely separated $\text{YI}=1$. An YI below 0 indicates that the classification rule for defining a positive test result must be reversed. It is worth mentioning that the YI is equivalent to the Kolmogorov--Smirnov measure of distance between the distributions of test outcomes in the diseased and nondiseased populations. Note that from Equation \ref{YIroc}, the YI can also be interpreted as the maximum vertical distance between the ROC curve and the chance diagonal. An appealing feature of the YI not present in the AUC is that it provides a criterion for choosing the threshold value to diagnose subjects in practice. The criterion is to choose the value $c^{*}$ that maximises Equation \ref{YIcdf} or $c^{*} = F_{\bar{D}}^{-1}(1-p^{*})$, with $p^{*}$ being the value that maximises Equation \ref{YIroc}. For further measures of diagnostic accuracy, such as partial areas under the ROC curve, where only a subset of FPFs or TPFs are considered, we refer the reader to \citet[][Chapter~4]{Pepe03}.

We finish this section highlighting that the ROC curve, as usually defined, measures the discriminatory capacity of a test under the particular classification rule that says that individuals with a test outcome larger than a pre-specified threshold are diseased, while those with a test outcome lower than the threshold are classified as nondiseased. The appropriateness of such classification rule relies on the standard assumption that larger test outcomes are more indicative of disease. However, this is not always the case. For instance, not only high but also low test results might be associated with disease. An example is provided in \cite{Martinez2017}. Therefore, one should be aware that the classification rule on which the usual definition of the ROC curve is based might not be the `optimal' one, in the sense that it might not be the classification rule based on $Y$ that provides the largest discriminatory capacity. We note that the optimality of the classification rule is directly related to the concavity of the resulting ROC curve and refer the reader to \cite{Fawcett2006}, \cite{Gneiting18} and \citet[][p.~71]{Pepe03} for a more extensive account on the importance of concave (also denoted in the literature as proper) ROC curves.

\subsection{ROC Curve and Related Indices Estimation}\label{rocest}
In what follows, let $\{y_{\bar{D}i}\}_{i=1}^{n_{\bar{D}}}$ and $\{y_{Dj}\}_{j=1}^{n_D}$ be two independent random samples of test outcomes from the nondiseased and diseased populations of size $n_{\bar{D}}$ and $n_D$, respectively. 

Statistical methods for estimating ROC curves have received wide attention in the literature. Plenty of parametric, semi, and nonparametric estimators have been proposed, both within frequentist and Bayesian paradigms. It would be an impossible task to cover, or even mention, all methods available. We succinctly describe the main idea behind each class of methods and further details can be found in the references provided. We give slightly more details about the nonparametric methods, as they are more widely applicable.

A fully parametric approach estimates the constituent distribution functions parametrically to arise at the induced ROC curve estimate. Let $F_D$ and $F_{\bar{D}}$ be parametrised in terms of $\theta_D$ and $\theta_{\bar{D}}$, respectively, i.e., $F_{D}(y)=F_{D}(y\mid\theta_D)$ and $F_{\bar{D}}(y)=F_{\bar{D}}(y\mid\theta_{\bar{D}})$. Estimating the parameters on the basis of test outcomes from each corresponding group, yields $\widehat{\theta}_D$ and $\widehat{\theta}_{\bar{D}}$, and the resultant ROC estimate is
\begin{equation*}
\widehat{\text{ROC}}(p)=1-F_{D}\{F_{\bar{D}}^{-1}(1-p\mid\widehat{\theta}_{\bar{D}})\mid\widehat{\theta}_D\}.
\end{equation*}
Typically, a normal distribution is assumed for both $F_D$ and $F_{\bar{D}}$, possibly after some transformation of the $Y_D$ and $Y_{\bar{D}}$ scales (e.g., the logarithmic one or a Box--Cox type of transformation). See \cite{Brownie86} and \cite{Goddard90} for examples of this approach.

In a semiparametric setting, the most common approach for ROC curve estimation is to assume a fully parametric form for the ROC curve, but making no assumptions about the distributions of the test outcomes themselves. These type of approaches have also been termed parametric distribution-free \citep{Pepe00, Alonzo02}. The most popular of these strategies is, perhaps, the binormal model, which postulates the existence of some unspecified strictly increasing transformation $H$, such that $H(Y_D)$ and $H(Y_{\bar{D}})$ follow a normal distribution. Specifically, and without loss of generality, if $H$ is such that $H(Y_D)\sim\text{N}(\mu,\sigma^2)$ and $H(Y_{\bar{D}})\sim\text{N}(0,1)$, then the binormal ROC model is written as 
\begin{equation*}
\text{ROC}(p)=\Phi\{a+b\Phi^{-1}(p)\}, \quad a=\frac{\mu}{\sigma},\quad b=\frac{1}{\sigma},
\end{equation*}
where $\Phi(\cdot)$ is the cumulative distribution function of the standard normal distribution. The appropriateness of the binormality assumption was discussed, among others, by \cite{Swets86} and \cite{Hanley96}, who concluded that it provides a good approximation to a large range of ROC curve shapes that occur in practice. Estimation of the binormal ROC curve reduces to the problem of estimating $a$ and $b$. The corresponding AUC has a closed-form expression given by $\Phi(a/\sqrt{1+b^2})$. Under the binormal model several estimation methods have been proposed. The earliest approach is due to \cite{Dorfman69}, but it was only applicable to ordinal test results; later \cite{Metz98} adapted it to the case of continuous test results by using a strategy that relies on categorising the outcomes into a finite number of categories and then applying the Dorfman and Alf procedure. \cite{Pepe00} and \cite{Alonzo02} suggest estimating the ROC curve by using procedures for fitting generalised linear models to binary data (these procedures will be further detailed in Section \ref{covariateroc}). \cite{Zou00} considered a method based on rank likelihood and \cite{Gu09} proposed a Bayesian approach that also uses a rank-based likelihood. We also mention the work of \cite{Cai04} who developed a profile maximum likelihood approach.

Apart from parametric and semiparametric approaches, several authors have also devoted their attention to the development of nonparametric methods, which are more generally applicable. All nonparametric methods reviewed here rely on (flexibly) estimating $F_D$ and $F_{\bar{D}}$ and plugging such estimates in Equation \ref{rocdef}. The most popular and simplest nonparametric method, due to \cite{Hsieh96}, is based on estimating $F_D$ and $F_{\bar{D}}$ by their corresponding empirical distribution functions, that is,
\begin{equation*} 
\widehat{F}_{D}(y)=\frac{1}{n_D}\sum_{j=1}^{n_D}I(y_{Dj}\leq y),\quad \widehat{F}_{\bar{D}}(y)=\frac{1}{n_{\bar{D}}}\sum_{i=1}^{n_{\bar{D}}}I(y_{\bar{D}i}\leq y).
\end{equation*}
Interestingly, the area under the empirical ROC curve is equal to the Mann--Whitney U statistic \citep{Bamber1975} 
\begin{equation*}
\widehat{\text{AUC}} = \frac{1}{n_D n_{\bar{D}}}\sum_{j = 1}^{n_D}\sum_{i = 1}^{n_{\bar{D}}}\left\{I\left(y_{Dj} > y_{\bar{D}i}\right)+\frac{1}{2}I\left(y_{Dj} = y_{\bar{D}i}\right)\right\}.
\end{equation*}
As it is clear from its definition, the empirical ROC curve is an increasing step function, which can be quite jagged, especially for small sample sizes and, as a consequence, might be unappealing in practice. To overcome the lack of smoothness of the empirical estimator, kernel-based methods for estimating the ROC curve have been developed. The earliest approach is due to \cite{Zou97}, who suggested estimating the density function in each population using kernel density estimates. Specifically,
\begin{equation*}
\widehat{f}_{D}(y)=\frac{1}{n_Dh_D}\sum_{j=1}^{n_D}k\left(\frac{y-y_{Dj}}{h_D}\right),\quad \widehat{f}_{\bar{D}}(y)=\frac{1}{n_{\bar{D}}h_{\bar{D}}}\sum_{i=1}^{n_{\bar{D}}}k\left(\frac{y-y_{\bar{D}i}}{h_{\bar{D}}}\right),
\end{equation*}
where $f_D$ ($f_{\bar{D}}$) corresponds to the density associated to $F_D$ ($F_{\bar{D}}$), $k(\cdot)$ is the kernel function, and $h_D$ ($h_{\bar{D}}$) is the bandwidth or smoothing parameter. The kernel considered was the biweight  and the corresponding distribution function estimates, $\widehat{F}_D$ and $\widehat{F}_{\bar{D}}$, were obtained by numerical integration. In a follow-up work, \cite{Zou98} suggested the use of the normal kernel and, in such case, the estimates of the distribution functions can be written as
\begin{equation*}
\widehat{F}_{D}(y)=\frac{1}{n_D}\sum_{j=1}^{n_D}\Phi\left(\frac{y-y_{Dj}}{h_D}\right), \quad \widehat{F}_{\bar{D}}(y)=\frac{1}{n_{\bar{D}}}\sum_{i=1}^{n_{\bar{D}}}\Phi\left(\frac{y-y_{\bar{D}i}}{h_{\bar{D}}}\right).
\end{equation*}
Still for the normal kernel, \cite{Lloyd98} has shown that the resulting estimate of the AUC has the following form
\begin{equation*}
\widehat{\text{AUC}}=\frac{1}{n_{D}n_{\bar{D}}}\sum_{j=1}^{n_D}\sum_{i=1}^{n_{\bar{D}}}\Phi\left(\frac{y_{Dj}-y_{\bar{D}i}}{\sqrt{h_D^2+h_{\bar{D}}^2}}\right).
\end{equation*}
The bandwidth, which controls the amount of smoothing and whose selection is critical to the performance of the estimator, was based on Silverman's rule of thumb \cite[][p.~48]{Silverman86}, which is optimal for data that are approximately bell-shaped distributed. Alternatively, the bandwidth can also be selected by least squares cross-validation; although this has not been proposed by the authors, it works quite well in practice for density estimation. The fact that the bandwidth proposed by \cite{Zou97} is not optimal for the ROC curve, because the latter depends on the distribution functions, and optimality for estimating density functions does not carry over the distribution functions, prompted \cite{Lloyd98} and \cite{Zhou02}, among other authors, to improve the above estimator by obtaining asymptotically optimal estimates for $F_D$ and $F_{\bar{D}}$. 

To finish this section, we turn our attention to Bayesian approaches and start with the nonparametric method of \cite{Erkanli06}, which models the distribution of test outcomes in each group via a Dirichlet process mixture of normal distributions \citep{Escobar95}, that is,
\begin{equation}\label{dpm}
F_{D}(y)=\int\Phi(y\mid\mu,\sigma^2)\text{d}G(\mu,\sigma^2),\quad G\sim\text{DP}(\alpha_D,G_{D}^{*}(\mu,\sigma^2)),
\end{equation}
with the distribution function in the nondiseased group following analogously. Here $G_D\sim\text{DP}(\alpha_D,G_{D}^{*})$ is used to denote that the mixing distribution $G_D$ follows a Dirichlet process (DP) prior \citep{Ferguson73} with centring distribution $G_D^{*}$, for which $E(G_D)=G_{D}^{*}$ and which  encapsulates any prior knowledge that might be known about $G_D$, and precision parameter $\alpha_D$, which controls the variability of $G_D$ around $G_D^{*}$. Larger values of $\alpha_D$ result in realisations $G_D$ that are closer to $G_D^{*}$. Unarguably, the most useful definition of the DP is its constructive definition due to \cite{Sethuraman94}, which postulates that $G_D$ can be written as
\begin{equation*}
G_D(\cdot)=\sum_{l=1}^{\infty}\omega_{Dl}\delta_{(\mu_{Dl},\sigma_{Dl}^2)}(\cdot),
\end{equation*}
where $\delta_a$ denotes a point mass at $a$, $(\mu_{Dl},\sigma_{Dl}^2)\overset{\text{iid}}\sim G_D^{*}(\mu,\sigma^2)$, and  the weights follow the so-called stick-breaking construction: $\omega_{D1}=v_{D1}$, $\omega_{Dl}=v_{Dl}\prod_{m<l}(1-v_{Dm})$, for $l\geq 2$, and $v_{Dl}\sim\text{Beta}(1,\alpha_D)$, for $l\geq 1$. Under Sethuraman's representation, the distribution function in Equation \ref{dpm} can be written as an infinite location-scale mixture of normal distributions, i.e.,
\begin{equation}\label{cdfsethu}
F_{D}(y)=\sum_{l=1}^{\infty}\omega_{Dl}\Phi(y\mid\mu_{Dl},\sigma_{Dl}^2).
\end{equation}
For the ease of posterior inference, a conjugate centring distribution is usually specified, i.e., $G_{D}^{*}\equiv\text{N}(\mu\mid m_D,S_D)\Gamma(\sigma^{-2}\mid a_D,b_D)$. A blocked Gibbs sampler \citep{Ishwaran02}, which relies on truncating the infinite mixture in Equation \ref{cdfsethu} to a finite number of components, say $L_{D}$, can then be used for conducting posterior inference, thus obtaining posterior samples of the weights, components' means and variances. Note that $L_D$ is not the number of components one expects to observe in the data but an upper bound on it. At iteration $s$ of the Gibbs sampler procedure, the ROC curve is computed as
\begin{align*}
\text{ROC}^{(s)}(p)&=1-F_{D}^{(s)}\{F_{\bar{D}}^{-1(s)}(1-p)\},\qquad s=1,\ldots,S,\\
F_{D}^{(s)}(y)&=\sum_{l=1}^{L_D}\omega_{Dl}^{(s)}\Phi(y\mid \mu_{Dl}^{(s)},\sigma_{Dl}^{2(s)}),\quad F_{\bar{D}}^{(s)}(y)=\sum_{l=1}^{L_{\bar{D}}}\omega_{\bar{D}l}^{(s)}\Phi(y\mid \mu_{\bar{D}l}^{(s)},\sigma_{\bar{D}l}^{2(s)}).
\end{align*}
As shown by the authors, the AUC admits the following closed-form expression
\begin{equation}\label{aucdpm}
\text{AUC}^{(s)}=\sum_{k=1}^{L_{\bar{D}}}\sum_{l=1}^{L_D}\omega_{\bar{D}k}^{(s)}\omega_{Dl}^{(s)}\Phi\left(\frac{a_{kl}^{(s)}}{\sqrt{1+b_{kl}^{2(s)}}}\right),\quad a_{kl}^{(s)}=\frac{\mu_{Dl}^{(s)}-\mu_{\bar{D}k}^{(s)}}{\sigma_{Dl}^{(s)}},\quad b_{kl}^{(s)}=\frac{\sigma_{\bar{D}k}^{(s)}}{\sigma_{Dl}^{(s)}}.
\end{equation}
At the end of the sampling procedure an ensemble composed of $S$ ROC curves/AUCs is available. The average of the ensemble is used as a point estimate and the variation in the ensemble is used to construct credible bands/intervals.

A somehow related approach is the Bayesian bootstrap (BB) ROC curve estimation procedure developed by \cite{Gu08}, which assumes that $F_D$ and $F_{\bar{D}}$ follow a Dirichlet process prior, rather than a Dirichlet process mixture as in the previous approach, i.e.,
\begin{align*}
\{y_{Dj}\}_{j=1}^{n_D},\mid F_D\sim F_D,\quad F_D\sim\text{DP}(\alpha_D,G_{D}^{*}),\\
\{y_{\bar{D}i}\}_{i=1}^{n_{\bar{D}}}\mid F_{\bar{D}}\sim F_{\bar{D}},\quad F_{\bar{D}}\sim\text{DP}(\alpha_{\bar{D}},G_{\bar{D}}^{*}),
\end{align*}
where by a slight abuse of notation we are also using here the same DP parameters'. From the conjugacy property of the DP \citep{Ferguson73}, which ensures that
\begin{equation}\label{dpconj}
F_D\mid\{y_{Dj}\}_{j=1}^{n_D}\sim\text{DP}\left(\alpha_D+n_D,\frac{\alpha_D}{\alpha_D+n_D}G_{D}^{*}+\frac{1}{\alpha_D+n_D}\sum_{j=1}^{n_D}\delta_{y_{Dj}}\right),
\end{equation}
it is clear that considering the noninformative limit of the DP, by letting $\alpha_D\rightarrow 0$ and  $\alpha_{\bar{D}}\rightarrow 0$, simplifies drastically the computational effort, as one does not even need to specify the centring distributions $G_{D}^{*}$ and $G_{\bar{D}}^{*}$ (an equivalent to Equation \ref{dpconj} holds for the nondiseased population). All that is needed is to generate from the uniform distribution over the simplex, which is equivalent to generating from a Dirichlet distribution with all parameters equal to one. The BB estimator of the ROC curve relies on a two-step procedure that makes use of the representation of the ROC curve as the distribution function of the diseased placement variable $U_D$. Specifically, as shown by the authors, it is only needed to 1) impute the variable $U_D=1-F_{\bar{D}}(Y_D)$ by plugging-in the survival function of $Y_{\bar{D}}$, generated from the BB resampling distribution given test outcomes $(y_{\bar{D1}},\ldots,y_{\bar{D}n_{\bar{D}}})$, and 2) compute the distribution function of $U_D$ based on the BB resample distribution to form one (of, say, $S$) realisation of the ROC curve. In fact, Step 1 is as simple as computing $U_{Dj}^{(s)}=\sum_{i=1}^{n_{\bar{D}}}q_{1i}^{(s)}I(y_{\bar{D}i}\geq y_{Dj})$, $j=1,\ldots,n_D$, and where $(q_{11}^{(s)},\ldots,q_{1n_{\bar{D}}}^{(s)})\sim\text{Dirichlet}(n_{\bar{D}};1,\ldots,1)$. In Step 2, we only need to calculate $\text{ROC}^{(s)}(p)=\sum_{j=1}^{n_D}q_{2j}^{(s)}I(U_{Dj}^{(s)}\leq p)$, with $(q_{21}^{(s)},\ldots,q_{2n_{D}}^{(s)})\sim\text{Dirichlet}(n_{D};1,\ldots,1)$. The AUC can also be expressed in closed form as $\text{AUC}^{(s)}=1-\sum_{j=1}^{n_D}q_{2j}^{(s)}U_{Dj}^{(s)}$.

Still within a Bayesian nonparametric framework, we mention the approach of \cite{Branscum08}, which is based on a different nonparametric prior, namely, a mixture of finite Polya trees, for modelling $F_D$ and $F_{\bar{D}}$. At last, for an overview article entirely dedicated to ROC curve estimation, we refer to \cite{Goncalves2014}.

Concerning the estimation of the Youden index and/or associated optimal threshold, for all approaches that rely on estimating the distribution functions of test outcomes, they can be obtained by simply plugging the corresponding estimates of $F_D$ and $F_{\bar{D}}$ in Equation \ref{YIcdf}. For the binormal model, where it is not assumed an explicit distribution for the test outcomes, Equation \ref{YIroc} should instead be used. For a detailed comparison among different methods (namely, empirical, kernel, and a pararametric one assuming normality on the original scale or after a Box--Cox transformation), we refer the reader to the article by \cite{Fluss05}.

\subsection{Illustration}
We now illustrate the methods described in the previous section with the HOMA-IR dataset. Recall that we seek to assess the accuracy of the HOMA-IR levels when predicting the presence of cardio-metabolic risk. Here we stratify the analysis by gender but disregard the age effect (i.e., HOMA-IR levels were pooled together regardless the age of the individuals). As we will be using both the kernel-based approach and the Dirichlet process mixture model with a normal kernel, the logarithm of HOMA-IR levels was considered. Figure 1 in the Supplementary Materials shows the estimated densities, by gender and in each population (individuals with and without cardio-metabolic risk), under the Dirichlet process mixture of normals model and the (normal) kernel method with bandwidth selected by Silverman's rule of thumb, and we can appreciate that both are very similar and follow the histograms of HOMA-IR levels quite closely. The estimated ROC curves using the four nonparametric methods described in the previous section are depicted in Figure \ref{pooledROC}. All methods produced very similar ROC curves. In Figure 2 of the Supplementary Materials we depict the same ROC curves but without the confidence/credible bands, so that the comparison between point estimates is clearer. The corresponding AUCs are reported in Table \ref{tabauc} and they are, both for men and women, close to $0.70$, revealing a mild accuracy of HOMA-IR levels for predicting cardio-metabolic risk. This comes as no surprise as Figure \ref{estdensities} already evidenced a quite considerable overlap of HOMA-IR levels in the two populations. Table 1 of the Supplementary Materials presents the Youden index and corresponding optimal HOMA-IR thresholds estimates that can be used to detect, in practice, individuals with higher cardio-metabolic risk.

\begin{figure}[h]
\includegraphics[width=14cm]{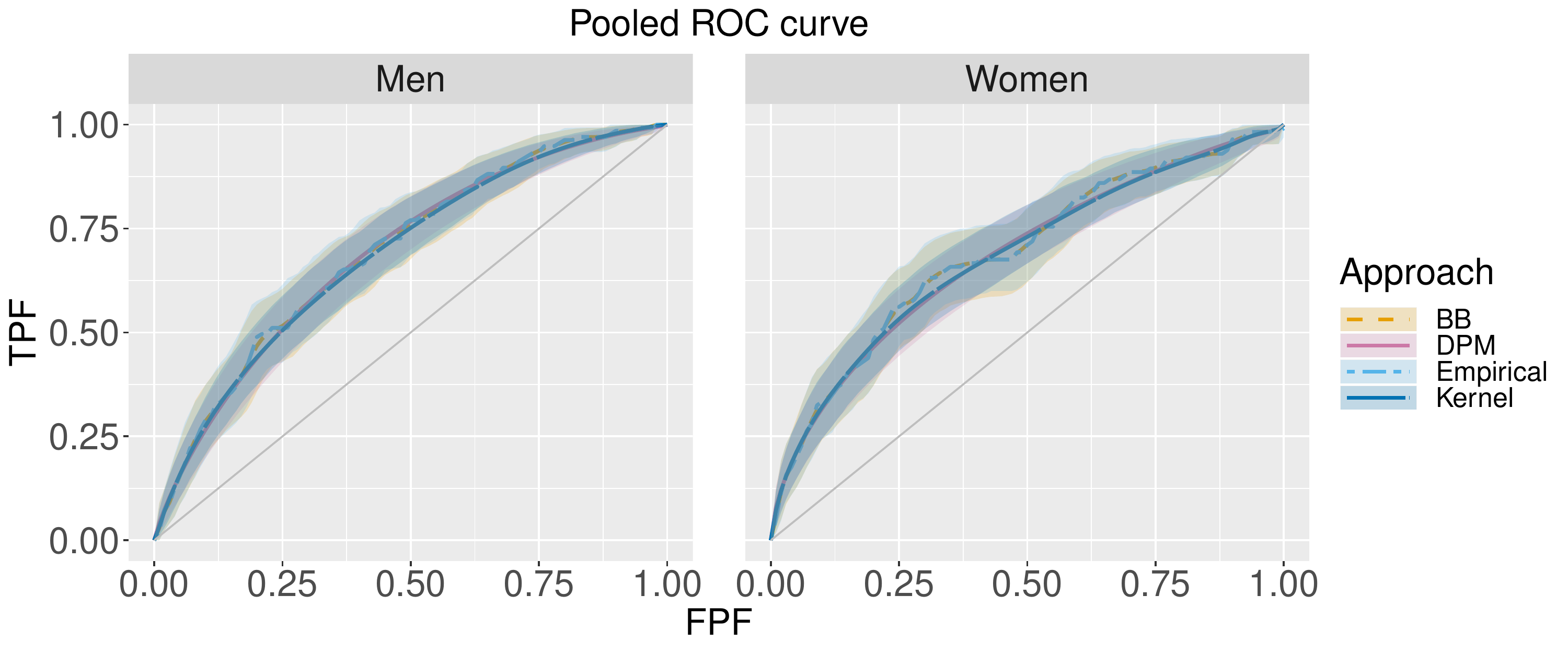}
\caption{Estimated ROC curves. The continuous lines correspond to point estimates and the shaded regions to the $95\%$ pointwise credible/confidence bands. Here BB stands for the Bayesian bootstrap method \citep{Gu08} and DPM for the Dirichlet process mixture of normals model \citep{Erkanli06}.}
\label{pooledROC}
\end{figure}

\begin{table}[h]
\tabcolsep7.5pt
\caption{AUC point estimates and $95\%$ credible/confidence intervals. Here BB stands for the Bayesian bootstrap method \citep{Gu08} and DPM for the Dirichlet process mixture of normals model \citep{Erkanli06}.}
\label{tabauc}
\begin{center}
\begin{tabular}{lccc}
& \multicolumn{2}{c}{AUC}\\
Approach & Women & Men \\
\hline
Empirical &  $0.691$ $(0.634, 0.736)$ &  $0.695$ $(0.647, 0.741)$\\
Kernel & $0.683$ $(0.629, 0.728)$  & $0.687$ $(0.641, 0.733)$\\
DPM & $0.685$ $(0.631, 0.736)$  & $0.691$ $(0.643, 0.736)$\\
BB & $0.691$ $(0.635, 0.743)$  & $0.695$ $(0.646, 0.740)$\\
\hline
\end{tabular}
\end{center}
\end{table}

\section{ROC CURVES AND COVARIATES}\label{covariateroc}

\subsection{Motivation}
\begin{figure}[h]
\includegraphics[width=14.5cm]{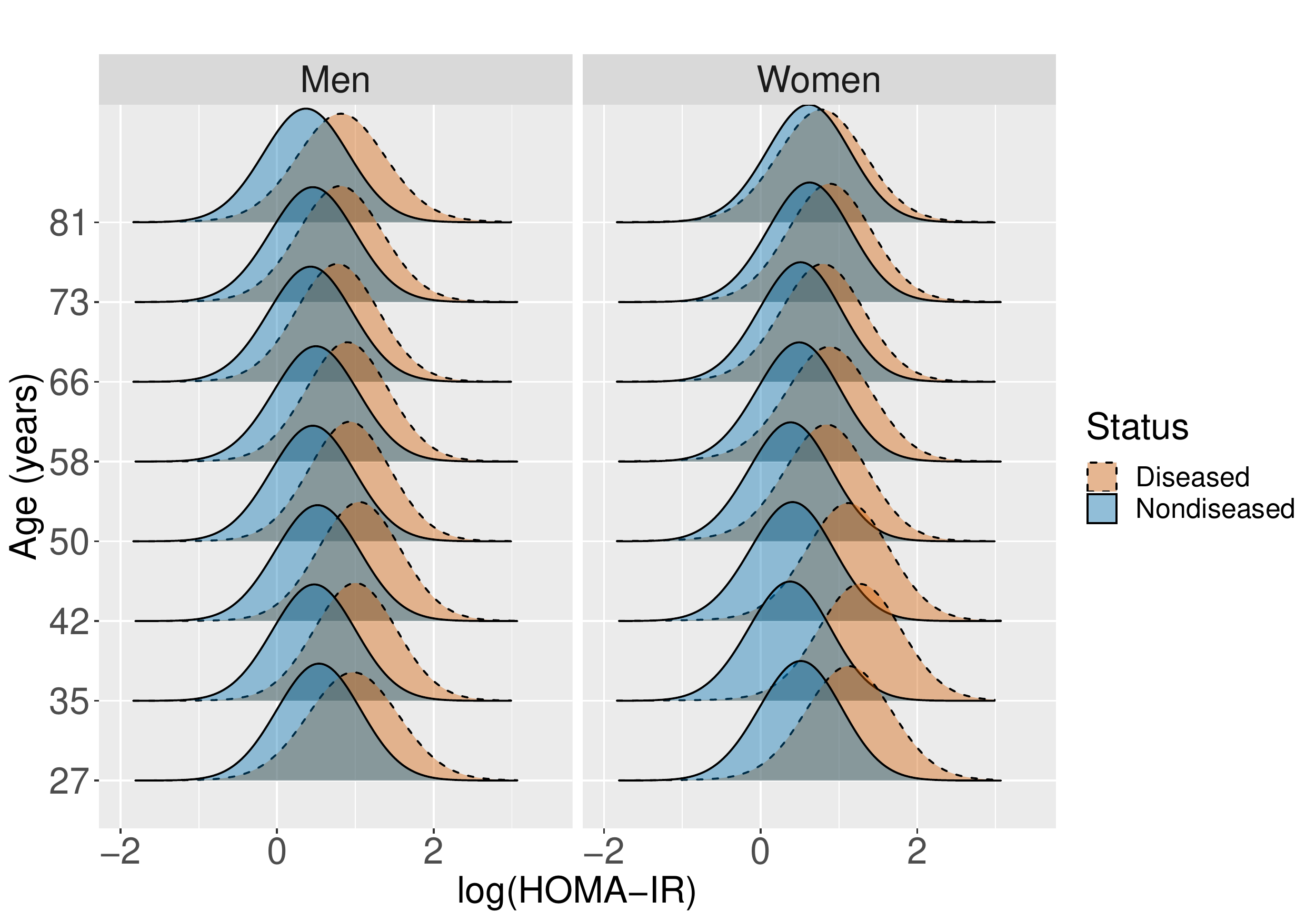}
\caption{Estimated density functions, obtained using a single-weights dependent Dirichlet process mixture of normals model, of $\log$ HOMA-IR levels in the diseased (dotted line, orange) and nondiseased (solid line, blue) populations, conditional on age and gender.}
\label{conditionaldensities}
\end{figure}

The definition of ROC curve given in Equation \ref{rocdef} implicitly assumes that both the diseased and nondiseased populations are homogeneous, at least, with regard to the performance of the test. However, this is rarely the case in practice. For instance, coming back to our motivating example, Figure \ref{conditionaldensities} shows the densities of $\log$ HOMA-IR levels conditional on the age and gender of the subjects. It can be noticed that, especially for women, the overlap between the two distributions of $\log$ HOMA-IR levels changes with age, and thus we expect the accuracy of $\log$ HOMA-IR levels to vary across age as well. This illustrates that, quite often, the distribution of test outcomes, either in the nondiseased or diseased population, or in both, is likely to vary with covariates. Examples of such covariates include subject-specific characteristics or different test settings. We note in passing that this does not necessarily mean covariates affecting the discriminatory capacity of the test. In particular, the distributions of test outcomes might experience a shift for different covariate values but their overlap might remain the same, case in which the accuracy of the test does not change, but still the thresholds used for defining a positive result will be covariate-specific (for further details we refer to \citealt[][Chapter~6]{Pepe03}, \citealt{Pardo2014}, and \citealt{Inacio18}).

\subsection{Notation and Definitions}
Let us now assume that along with $Y_D$ and $Y_{\bar{D}}$, covariate vectors $\mathbf{X}_{D}$ and $\mathbf{X}_{\bar{D}}$ are also available. For ease of notation, we assume that the covariates of interest are the same in both populations, although this is not always necessarily the case (e.g., disease stage is, obviously, a disease-specific covariate).

As a natural extension of the ROC curve, the conditional or covariate-specific ROC curve, given a covariate value $\mathbf{x}$, is defined as
\begin{equation}\label{roccov}
\text{ROC}(p\mid \mathbf{x})=1-F_{D}\{F_{\bar{D}}^{-1}(1-p\mid\mathbf{x})\mid\mathbf{x}\},\quad 0\leq p \leq 1,
\end{equation}
where $F_{D}(y\mid\mathbf{x})=\Pr(Y_D\leq y\mid\mathbf{X}_{D}=\mathbf{x})$ denotes the conditional distribution function in the diseased group, with $F_{\bar{D}}(y\mid\mathbf{x})$ being defined similarly. The covariate-specific counterparts of the AUC and YI are given by
\begin{align}
\text{AUC}(\mathbf{x})&=\int_{0}^{1}\text{ROC}(p\mid\mathbf{x})\text{d}p \label{covauc} \\
 \text{YI}(\mathbf{x}) &=  \max_{c} \{ F_{\bar{D}}(c\mid\mathbf{x}) -F_{D}(c\mid\mathbf{x})\} \label{covyidf}\\
 &=\max_p\{\text{ROC}(p\mid\mathbf{x})-p\} \label{covyiroc}
\end{align}
For each value of $\mathbf{x}$ we might obtain a different ROC curve (AUC and/or Youden index) and, therefore, also a
possible different accuracy. Understanding the influence of covariates on the accuracy of a diagnostic test will help in determining the optimal and suboptimal populations where to perform the diagnostic tests on.

\subsection{Covariate-specific ROC curve estimation}\label{covroc}
Approaches to estimation of the covariate-specific ROC curve can be broadly divided in two categories \citep{Pepe98}. Induced methodologies model the distribution of test outcomes in the diseased and nondiseased populations separately and then compute the induced ROC curve. On the other hand, direct methodologies assume a regression model directly on the covariate-specific ROC curve. In what follows, we now let $\{(\mathbf{x}_{\bar{D}i},y_{\bar{D}i})\}_{i=1}^{n_{\bar{D}}}$ and $\{(\mathbf{x}_{Dj},y_{Dj})\}_{j=1}^{n_D}$ be two independent random samples of covariates and test outcomes from the nondiseased and diseased populations of size $n_{\bar{D}}$ and $n_D$, respectively. Further, for all $i = 1,\ldots,n_{\bar{D}}$ and $j = 1,\ldots,n_D$, let $\mathbf{x}_{\bar{D}i}=(x_{\bar{D}i,1},\ldots, x_{\bar{D}i,q})^{\prime}$ and $\mathbf{x}_{Dj}=(x_{Dj,1},\ldots, x_{Dj,q})^{\prime}$ be $q-$dimensional vectors of covariates, which can be either continuous or categorical.

\subsubsection{Induced methodology}
For clarity in the presentation, within the induced methodology, we distinguish between two types of approaches. Both aim at estimating the constituent components of the covariate-specific ROC curve, i.e., the conditional distribution of test results in the diseased and nondiseased populations (see Equation \ref{roccov}). However, whereas the first set of methods do it through the specification of a location-scale regression model for the test outcomes in each population, the second set focus on directly modelling the conditional distributions.

We start by presenting the first mentioned induced approach. Specifically, the relationship between covariates and test outcomes in each population is given by location-scale regression models
\begin{equation}\label{locationscale}
Y_{D}=\mu_{D}(\mathbf{X}_D)+\sigma_{D}(\mathbf{X}_D)\varepsilon_{D},\qquad Y_{\bar{D}}=\mu_{\bar{D}}(\mathbf{X}_{\bar{D}})+\sigma_{\bar{D}}(\mathbf{X}_{\bar{D}})\varepsilon_{\bar{D}},
\end{equation}
where $\mu_{D}(\mathbf{x})=E(Y_D\mid\mathbf{X}_D=\mathbf{x})$ and $\sigma_{D}^{2}=\text{var}(Y_D\mid\mathbf{X}_D=\mathbf{x})$ are, respectively, the conditional mean and variance of $Y_D$ given $\mathbf{X}_D=\mathbf{x}$, with $\mu_{\bar{D}}$ and $\sigma_{\bar{D}}^{2}$  being analogously defined. The error terms $\varepsilon_{D}$ and $\varepsilon_{\bar{D}}$ are assumed to be independent of each other and of the covariates, with zero mean, unit variance, and distribution function $F_{\varepsilon_{D}}$ and $F_{\varepsilon_{\bar{D}}}$, respectively. Given the independence between the error and the covariates in the location-scale regression models in Equation \ref{locationscale}, it is easy to show that
\begin{equation*}
F_{D}(y\mid\mathbf{x})=F_{\varepsilon_{D}}\left(\frac{y-\mu_{D}(\mathbf{x})}{\sigma_D(\mathbf{x})}\right),\qquad F_{\bar{D}}(y\mid\mathbf{x})=F_{\varepsilon_{\bar{D}}}\left(\frac{y-\mu_{\bar{D}}(\mathbf{x})}{\sigma_{\bar{D}}(\mathbf{x})}\right).
\end{equation*}
An analogous relationship can be established between the conditional quantile function of test outcomes given the covariates and the quantile function of the error terms, namely
\begin{equation*}
F_{D}^{-1}(p\mid\mathbf{x})=\mu_{D}(\mathbf{x})+\sigma_{D}(\mathbf{x})F_{\varepsilon_{D}}^{-1}(p),\quad F_{\bar{D}}^{-1}(p\mid\mathbf{x})=\mu_{\bar{D}}(\mathbf{x})+\sigma_{\bar{D}}(\mathbf{x})F_{\varepsilon_{\bar{D}}}^{-1}(p).
\end{equation*}
The covariate-specific ROC curve, for a given covariate value $\mathbf{x}$, can therefore be expressed as
\begin{align*}
\text{ROC}(p\mid\mathbf{x})=1-F_{\varepsilon_{D}}\left\{\frac{\mu_{\bar{D}}(\mathbf{x})-\mu_{D}(\mathbf{x})}{\sigma_D(\mathbf{x})}+\frac{\sigma_{\bar{D}}(\mathbf{x})}{\sigma_{D}(\mathbf{x})}F_{\varepsilon_{\bar{D}}}^{-1}(1-p)\right\},\quad 0\leq p \leq 1.
\end{align*}
This formulation allows expressing the covariate-specific ROC curve in terms of the distribution and quantile functions of the regression errors, which are not conditional, thus reducing the computational burden.

Thus far we have described this form of induced ROC methodology in its most general form. Particular cases have been addressed in the literature. In particular, \cite{Faraggi03} assumed a normal linear homoscedastic  model in each population, that is
\begin{equation*}
\mu_{D}(\mathbf{x})=\tilde{\mathbf{x}}^{\prime}\boldsymbol{\beta}_D, \quad \sigma_D(\mathbf{x})=\sigma_D,\quad F_{\varepsilon_{D}}(\cdot)=\Phi(\cdot),
\end{equation*}
with $\tilde{\mathbf{x}}^{\prime}=(1,\mathbf{x}^{\prime})$ and $\boldsymbol{\beta}_D=(\beta_{D0},\ldots,\beta_{Dq})^{\prime}$ is a $(q+1)-$dimensional vector of (unknown) regression coefficients. All quantities are analogously defined for the nondiseased population. Estimates of the regression coefficients $\boldsymbol{\beta}_{D}$ and $\boldsymbol{\beta}_{\bar{D}}$ are obtained by ordinary least squares on the basis of the samples $\{(\mathbf{x}_{Dj},y_{Dj})\}_{j=1}^{n_D}$ and $\{(\mathbf{x}_{\bar{D}i},y_{\bar{D}i})\}_{i=1}^{n_{\bar{D}}}$, respectively. The variances are then straightforwardly estimated as
\begin{equation*} 
\widehat{\sigma}_{D}^2=\frac{\sum_{j=1}^{n_D}(y_{Dj}-\tilde{\mathbf{x}}_{Dj}^{\prime}\widehat{\boldsymbol{\beta}}_D)^2}{n_D-q-1},\qquad \widehat{\sigma}_{\bar{D}}^2=\frac{\sum_{i=1}^{n_{\bar{D}}}(y_{\bar{D}i}-\tilde{\mathbf{x}}_{\bar{D}i}^{\prime}\widehat{\boldsymbol{\beta}}_{\bar{D}})^2}{n_{\bar{D}}-q-1}.
\end{equation*}
The corresponding covariate-specific ROC curve is given by
\begin{equation*}
\widehat{\text{ROC}}(p\mid\mathbf{x})=1-\Phi\{a(\mathbf{x})+b\Phi^{-1}(1-p)\},\quad a(\mathbf{x})=\tilde{\mathbf{x}}^{\prime}\frac{(\widehat{\boldsymbol{\beta}}_{\bar{D}}-\widehat{\boldsymbol{\beta}}_D)}{\widehat{\sigma}_D},\quad b=\frac{\widehat{\sigma}_{\bar{D}}}{\widehat{\sigma}_D}.
\end{equation*}
As for the binormal ROC curve in the no-covariate case, the AUC under this model is given by $\Phi(-a(\mathbf{x})/\sqrt{1+b^2})$.

Alternatively, and less restrictive, \cite{Pepe98} suggests to estimate the distribution function of the errors in each population by
the corresponding empirical distribution function of the estimated standardised residuals. Note that in the original paper the same distribution was assumed in both populations, but we are presenting here the more general case in which each population has its own distribution, i.e., 
\begin{equation*}
\widehat{F}_{\varepsilon_{D}}(y)=\frac{1}{n_D}\sum_{j=1}^{n_D}I(\widehat{\varepsilon}_{Dj}\leq y),\qquad \widehat{\varepsilon}_{Dj}=\frac{y_{Dj}-\tilde{\mathbf{x}}_{Dj}^{\prime}\widehat{\boldsymbol{\beta}}_D}{\widehat{\sigma}_D}, 
\end{equation*}
with $\widehat{F}_{\varepsilon_{\bar{D}}}(y)$ and $\widehat{\varepsilon}_{\bar{D}i}$, $i=1,\ldots,n_{\bar{D}}$, are defined in a similar fashion. The covariate-specific ROC curve is finally computed in an analogous way as
for the method of \cite{Faraggi03} as
\begin{equation*}
\widehat{\text{ROC}}(p\mid\mathbf{x})=1-\widehat{F}_{\varepsilon_{D}}\{a(\mathbf{x})+b\widehat{F}_{\varepsilon_{\bar{D}}}^{-1}(1-p)\},\qquad 0\leq p\leq 1. 
\end{equation*}
The covariate-specific AUC also admits a closed-form expression which can be regarded as a covariate-specific Mann--Whitney type of statistic, that is, 
\begin{equation}\label{closedauc}
\widehat{\text{AUC}}(\mathbf{x})=\frac{1}{n_{D}n_{\bar{D}}}\sum_{j=1}^{n_D}\sum_{i=1}^{n_{\bar{D}}}I\{\widehat{\mu}_{\bar{D}}(\mathbf{x})+\widehat{\sigma}_{\bar{D}}\widehat{\varepsilon}_{\bar{D}i}\leq \widehat{\mu}_{D}(\mathbf{x})+\widehat{\sigma}_{D}\widehat{\varepsilon}_{Dj}\}.
\end{equation}
Still in a semiparametric context, \cite{Zheng04} proposed an estimator for the covariate-specific ROC curve in which the distribution of the error terms is unknown and allowed to depend on covariates (and so, strictly speaking, the underlying models for the test outcomes are no longer location-scale regression models) but, as in the previous two approaches, the effect of the covariates on the conditional means and variances is modelled parametrically. In a Bayesian context,  \cite{Rodriguez14} proposed a semiparametric model, where the (marginal) error terms are assumed to follow a Student-$t$ distribution and the conditional mean and variance functions are modelled nonparametrically through Gaussian process priors.

Within a nonparametric frequentist perspective, \cite{Yao10}, \cite{Gonzalez11}, and \cite{Rodriguez11} all proposed a kernel-based approach to estimate the mean and variance functions in Equation \ref{locationscale} but, as proposed by these authors, the method can only deal with one continuous covariate. Both the regression and the variance functions are estimated using local polynomial kernel smoothers \citep{Fan96}. Estimation proceeds in a sequential manner: 1) the regression functions in the diseased and nondiseased populations are estimated first on the basis of $\{(x_{Dj},y_{Dj})\}_{j=1}^{n_{D}}$ and $\{(x_{\bar{D}i},y_{\bar{D}i})\}_{i=1}^{n_{\bar{D}}}$, respectively, and 2) the variance function is estimated next on the basis of the samples $\{(x_{Dj},[y_{Dj}-\widehat{\mu}_{D}(x_{Dj})]^2)\}_{j=1}^{n_{D}}$ and $\{(x_{\bar{D}i},[y_{\bar{D}i}-\widehat{\mu}_{\bar{D}}(x_{\bar{D}i})]^2)\}_{i=1}^{n_{\bar{D}}}$. Both steps involve the selection of a smoothing parameter and that can be done, for instance, via least squares cross-validation. Once estimates of the mean and variance functions are available, the standardised residuals can be calculated and, as in Pepe's method, their empirical distribution function is used to estimate the distribution of the error terms. The covariate-specific AUC can also be written in the form of Equation \ref{closedauc}, with the mean and variance functions replaced by their corresponding kernel-based counterparts. Because the estimator of the conditional ROC curve is based on the emprirical distribution function (of the standardised residuals), the resulting estimator is not smooth and, in order to overcome this drawback, \cite{Gonzalez11} also proposed an estimator that makes use of a further bandwidth and does the convolution with a continuous kernel, namely
\begin{equation*}
\widehat{\text{ROC}}_{h}(p\mid\mathbf{x})=1-\int\widehat{F}_{\varepsilon_{D}}\left(a(x)+\widehat{F}_{\varepsilon_{\bar{D}}}^{-1}(1-p+hu)b(x)\right)k(u)\text{d}u,
\end{equation*}
where $a(x)=\frac{\widehat{\mu}_{\bar{D}}(x)-\widehat{\mu}_{D}(x)}{\widehat{\sigma}_{D}(x)}$, $b(x)=\frac{\widehat{\sigma}_{\bar{D}}(x)}{\widehat{\sigma}_{D}(x)}$, and $k(\cdot)$ is a kernel function. Note that when $h=0$ the non-smooth estimator is recovered.

We now briefly detail the approach of \cite{Inacio13} which, by opposition to the previous approaches, is based on directly modelling the conditional distribution function of test outcomes in the diseased and nondiseased populations, allowing it to smoothly change as a function of the covariates. Specifically, the authors use a single-weights linear dependent Dirichlet process mixture of normals to model the conditional distribution in each population
\begin{equation*}
F_{D}(y\mid\mathbf{x})=\int \Phi(y\mid\mu(\mathbf{x},\boldsymbol{\beta}),\sigma^2)\text{d}G_{D}(\boldsymbol{\beta},\sigma^2),\quad G_D\sim\text{DP}(\alpha_D,G_{D}^{*}(\boldsymbol{\beta},\sigma^2)),
\end{equation*}
with the conditional distribution function in the nondiseased population following in an analogous manner. This model can be regarded as an extension to the conditional case of the method of \cite{Erkanli06}. As in the no-covariate case, using Sethuraman's representation, the conditional distribution can be expressed as
\begin{equation}\label{ddp}
 F_{D}(y\mid\mathbf{x})=\sum_{l=1}^{\infty}\omega_{Dl}\Phi\left(y\mid\mu(\mathbf{x},\boldsymbol{\beta}_{Dl}),\sigma_{Dl}^2\right),
\end{equation}
 with the weights matching those from the stick-breaking construction as specified in Erkanli's model. Notice that the only difference between Equations \ref{cdfsethu} and \ref{ddp} is that now the mean of each component depends on covariates. Regarding the specification of the components' means, it has been recommended (see \citealt{Inacio18} for more details) to use a flexible formulation, so that a large number of (conditional) density shapes' are well-approximated. In particular, cubic B-splines basis functions are used for continuous covariates and, as a result, we write
 \begin{equation*}
\mu(\mathbf{x},\boldsymbol{\beta}_{Dl})=\mathbf{z}_{D}^{\prime}\boldsymbol{\beta}_{Dl}, \quad l\geq 1,\quad j =1,\ldots,n_D,
\end{equation*}
where $\mathbf{z}_{D}$ is the vector containing the intercept, the cubic B-splines basis representation of the continuous covariates, the categorical covariates (if any), and their interaction(s) with the smoothed continuous covariate(s) (if believed to exist). Also, $\boldsymbol{\beta}_{Dl}$ collects, for the $l$th component, the regression coefficients associated with the aforementioned covariate vector. The regression coefficients and variances associated with each component are sampled from a conjugate centring distribution $(\boldsymbol{\beta}_{Dl},\sigma_{Dl}^{-2})\overset{\text{iid}}\sim\text{N}(\mathbf{m}_D,\mathbf{S}_D)\Gamma(a_D,b_D)$ and, as in the unconditional case, the blocked Gibbs sampler is used to simulate draws from the posterior distribution. At iteration $s$ of the Gibbs sampler procedure, the covariate-specific ROC curve is computed as
\begin{align*}
\text{ROC}^{(s)}(p\mid\mathbf{x})&=1-F_{D}^{(s)}\{F_{\bar{D}}^{-1(s)}(1-p\mid\mathbf{x})\mid\mathbf{x}\},\qquad s=1,\ldots,S,\\
F_{D}^{(s)}(y\mid\mathbf{x})&=\sum_{l=1}^{L_D}\omega_{Dl}^{(s)}\Phi(y\mid \mathbf{z}_{D}^{\prime}\boldsymbol{\beta}_{Dl}^{(s)},\sigma_{Dl}^{2(s)}),\quad F_{\bar{D}}^{(s)}(y\mid\mathbf{x})=\sum_{l=1}^{L_{\bar{D}}}\omega_{\bar{D}l}^{(s)}\Phi(y\mid \mathbf{z}_{\bar{D}}^{\prime}\boldsymbol{\beta}_{\bar{D}l}^{(s)},\sigma_{\bar{D}l}^{2(s)}).
\end{align*}
The covariate-specific AUC admits exactly the same closed form expression as in Equation \ref{aucdpm}, with the obvious difference that the components' means are covariate-dependent, i.e., we now have
\[
a_{kl}^{(s)}(\mathbf{x}) = \frac{\mu(\mathbf{x},\boldsymbol{\beta}_{Dl}^{(s)}) - \mu(\mathbf{x},\boldsymbol{\beta}_{\bar{D}k}^{(s)})}{\sigma_{Dl}^{(s)}}.
\]
Point and interval estimates for the covariate-specific ROC curve and AUC can be obtained from the corresponding ensembles of posterior realisations.

Another estimator for the conditional ROC curve that also directly models the conditional distribution of test outcomes, but based on kernel methods, was proposed by \cite{Lopez08}.

In what concerns estimation of the covariate-specific Youden index and/or associated threshold, because all induced approaches, in a more or less direct way, provide an estimate of the conditional distribution function of the test outcomes in each population, these can be plugged in the definition in Equation \ref{covyidf}, so that estimates of these quantities can be obtained. We also mention here the work by \cite{Xu14}, where the authors propose an approach that directly estimates the covariate-specific YI and threshold value without the need of first estimating the conditional distribution functions.

\subsubsection{Direct methodology}
In contrast to the induced approach, in the direct methodology the effect of the covariates is directly evaluated on the ROC curve, with its general form given by the following regression model
\begin{equation}\label{rocglm}
\text{ROC}(p\mid\mathbf{x})=g\{\mu(\mathbf{x})+h_0(p)\},\qquad 0\leq p \leq 1,
\end{equation}
where $\mu(\mathbf{x})$ collects the effects of the covariates on the ROC curve, $h_0(p)$ is an unknown monotonic increasing function of the FPF related to the shape of the ROC curve and $g$ is the inverse of the link function. Unlike in standard regression analysis, the response variable of the model presented in Equation \ref{rocglm} is not directly observable. However, note that the covariate-specific ROC curve can be  re-expressed as
\begin{align}
\text{ROC}(p\mid\mathbf{x})&=1-F_{D}\{F_{\bar{D}}^{-1}(1-p\mid\mathbf{x})\mid\mathbf{x}\} \nonumber\\
&=1-\Pr\{Y_D\leq F_{\bar{D}}^{-1}(1-p\mid\mathbf{x})\mid\mathbf{X}_D=\mathbf{x}\}\nonumber\\
&=\Pr\{1-F_{\bar{D}}(Y_D\mid\mathbf{x})<p\mid\mathbf{X}_D=\mathbf{x}\} \nonumber\\
&=E[I(1-F_{\bar{D}}(Y_D\mid\mathbf{x})<p)\mid\mathbf{X}_D=\mathbf{x}],\label{rocdpv}
\end{align}
and, in particular, as highlighted by Equation \ref{rocdpv}, it can be interpreted as the conditional expectation of the binary variable 
$I(1-F_{\bar{D}}(Y_D\mid\mathbf{x})<p)$ and, therefore, the ROC regression model in Equation \ref{rocglm} can be viewed as a regression model for $I(1-F_{\bar{D}}(Y_D\mid\mathbf{x})<p)$. Note that $1-F_{\bar{D}}(Y_D\mid\mathbf{X}_{D}=\mathbf{x})$ is nothing more than the conditional diseased placement value, that is, a covariate-specific version of the $U_D$ variable introduced in Section \ref{acc_measures}.

Different estimation proposals, which differ in the assumptions made about $g$, $\mu$, and $h_0$, have been suggested in the literature. In \cite{Pepe00} and \cite{Alonzo02}, $g$ is assumed to be known (e.g., $g(\cdot)=\Phi(\cdot)$), the effect of the covariates on the conditional ROC curve is assumed to be linear, i.e., $\mu(\mathbf{x})=\mathbf{x}^{\prime}\boldsymbol{\beta}$, and the baseline function $h_0$ is assumed to have a parametric form given by $h_{0}(p)=\sum_{k=1}^{K}\alpha_k h_k(p)$, where $\boldsymbol{\alpha}=(\alpha_1,\ldots,\alpha_K)^{\prime}$ is a vector of unknown parameters and $h(p)=(h_1(p),\ldots,h_K(p))$ are known functions. Note that the binormal model for the (unconditional) ROC curve arises when no covariates are considered and for $g(\cdot)=\Phi(\cdot)$, $h_1(p)=1$, and $h_2(p)=\Phi^{-1}(p)$. \cite{Cai02} and \cite{Cai04glm} studied a more flexible model by leaving $h_0$ completely unspecified, but the function $\mu$ is still modelled in a linear way and $g$ is also considered to be known. In general, models like those in Equation \ref{rocglm} with parametric specifications for $\mu$ define the so-called class of ROC-GLMs due to the similarities with generalised linear models \citep{Pepe00}. In contrast to the previous cited works, \cite{Lin12} developed a semiparametric model where both the link and baseline functions are completely unknown and $\mu$ is assumed to have a parametric form. Finally, \cite{Rodriguez11new} assumes that $g$ is known but an additive smooth structure is assumed for $\mu(\mathbf{x})$, i.e., $\mu(\mathbf{x})=\beta+\sum_{k=1}^{q}f_k(x_k)$, where $f_1,\ldots,f_q$ are unknown nonparametric functions and $h_0$ also remains unspecified.

Regardless of whether the specification in Equation \ref{rocglm} involves a generalised linear or additive model structure, the estimation process is similar and can be described as given in the following steps. First, one must choose a set of FPFs, say $0\leq p_l\leq 1$ for $l=1,\ldots,n_P$, where the covariate-specific ROC curve will be evaluated. Second, an estimate of $F_{\bar{D}}(\cdot\mid\mathbf{x})$, say $\widehat{F}_{\bar{D}}(\cdot\mid\mathbf{x})$, on the basis of the sample $\{(\mathbf{x}_{\bar{D}i},y_{\bar{D}i})\}_{i=1}^{n_{\bar{D}}}$, must be obtained. Third, one should calculate the estimated placement value for each disease observation $1-\widehat{F}_{\bar{D}}(y_{Dj}\mid\mathbf{x}_{Dj})$, for $j=1,\ldots,n_D$. The fourth step involves the calculation of the binary indicators $I(1-
\widehat{F}_{\bar{D}}(y_{Dj}\mid\mathbf{x}_{Dj})\leq p_l)$, for $j=1,\ldots,n_D$ and $l=1,\ldots,n_P$. Lastly, in fifth, the model $g(\mu(\mathbf{x})+h_0(p))$ is fitted as a regression model for binary data with indicators $I(1-\widehat{F}_{\bar{D}}(y_{Dj}\mid\mathbf{x}_{Dj})\leq p_l)$ as the response variable and covariates $\mathbf{x}_{Dj}$ and $h(p_l)$ (when $h_0$ is modelled parametrically) or $p_l$ (when $h_0$ is left unspecified), for $j=1,\ldots,n_D$ and $l=1,\ldots,n_P$. We note that the above algorithm does not apply for the estimation of the proposals described in \cite{Cai02}, \cite{Cai04glm}, and \cite{Lin12}. For conciseness we do not present here the details of their approaches, but refer the readers to the respective articles.

Regarding the estimation of the conditional AUC within the direct methodology, the obvious way is to simply plug-in an estimate for the conditional ROC curve in Equation \ref{covauc}, and approximate the integral using numerical integration methods. However, this approach might not be the most efficient one, and several methods to \textit{directly} estimate $\text{AUC}(\mathbf{x})$ have been proposed in the literature. We mention here the articles by \cite{Dodd03a,Dodd03b} and \cite{Cai08}, where semiparametric regression models for the conditional (partial) AUC are proposed. For the Youden index (and associated threshold value), to the best of our knowledge, no \textit{direct} estimators have been proposed. Estimation, in this case, requires making use of Equation \ref{covyiroc}, with $\text{ROC}(p\mid\mathbf{x})$ being replaced by its estimate. Note that, once we obtain the (conditional) FPF at which the maximum of \ref{covyiroc} is attained, an estimate of the associated conditional threshold value can be obtained using the estimator of $F_{\bar{D}}(\cdot\mid\mathbf{x})$ needed in the second step of the above described algorithm.

\subsection{Covariate-adjusted ROC curve}
The covariate-specific ROC curve and associated AUC and YI assess the accuracy of the test for specific covariate values. It would, however, be useful to have a global summary measure that also takes covariate information into account. 
The covariate-adjusted ROC (AROC) curve proposed by \cite{Janes09} is exactly one of such measures. It is defined as
\begin{equation*}
\text{AROC}(p)=\int \text{ROC}(p\mid\mathbf{x})\text{d}H_{D}(\mathbf{x}),
\end{equation*}
where $H_{D}(\mathbf{x}) = \Pr(\mathbf{X}_{D}\leq \mathbf{x})$ is the distribution function of $\mathbf{X}_{D}$. That is, the AROC curve is a weighted average of covariate-specific ROC curves, weighted according to the distribution of the covariates in the diseased group. As shown by the authors, the AROC curve can also be expressed as
\begin{align*}
\text{AROC}(p) & = \Pr\{Y_{D}>F_{\bar{D}}^{-1}(1-p\mid \mathbf{X}_{D})\} =\Pr\{1-F_{\bar{D}}(Y_D\mid\mathbf{X}_{D})\leq p\},
 \end{align*}
emphasising that the AROC curve at a FPF of $p$ is the overall TPF when the thresholds used for defining a positive test result are covariate-specific and chosen to ensure that the FPF is $p$ in each subpopulation defined by the covariate values. We refer to \cite{Janes09}, \cite{Rodriguez11}, and \cite{Inacio18} for the different estimation methods available for the ROC curve. 

A natural question to ask is when to use the covariate-specific ROC curve and the covariate-adjusted ROC curve. Very briefly, and without going into details, when the accuracy of the test does change with the covariates (i.e., when the separation between the distributions of test outcomes changes for different covariate levels), the covariate-specific ROC curve should be the primary tool to be used. On the other hand, if the distributions of the test outcomes change with covariates but not the accuracy of the test (i.e., if the overlap between the distributions of test outcomes remains the same for different covariate levels), then the covariate-adjusted ROC curve, which in this case corresponds to the common covariate-specific ROC curve, should be instead reported. For a lengthy discussion of this point, see \citet[][Chapter~6]{Pepe03}, \cite{Janes08a}, \cite{Janes08b}, and \cite{Inacio18}. Also, for a recent overview article focusing exclusively on ROC curves and covariates, we refer to \cite{Pardo2014}.

\subsection{Illustration}
\begin{figure}[h]
\includegraphics[width=14.5cm]{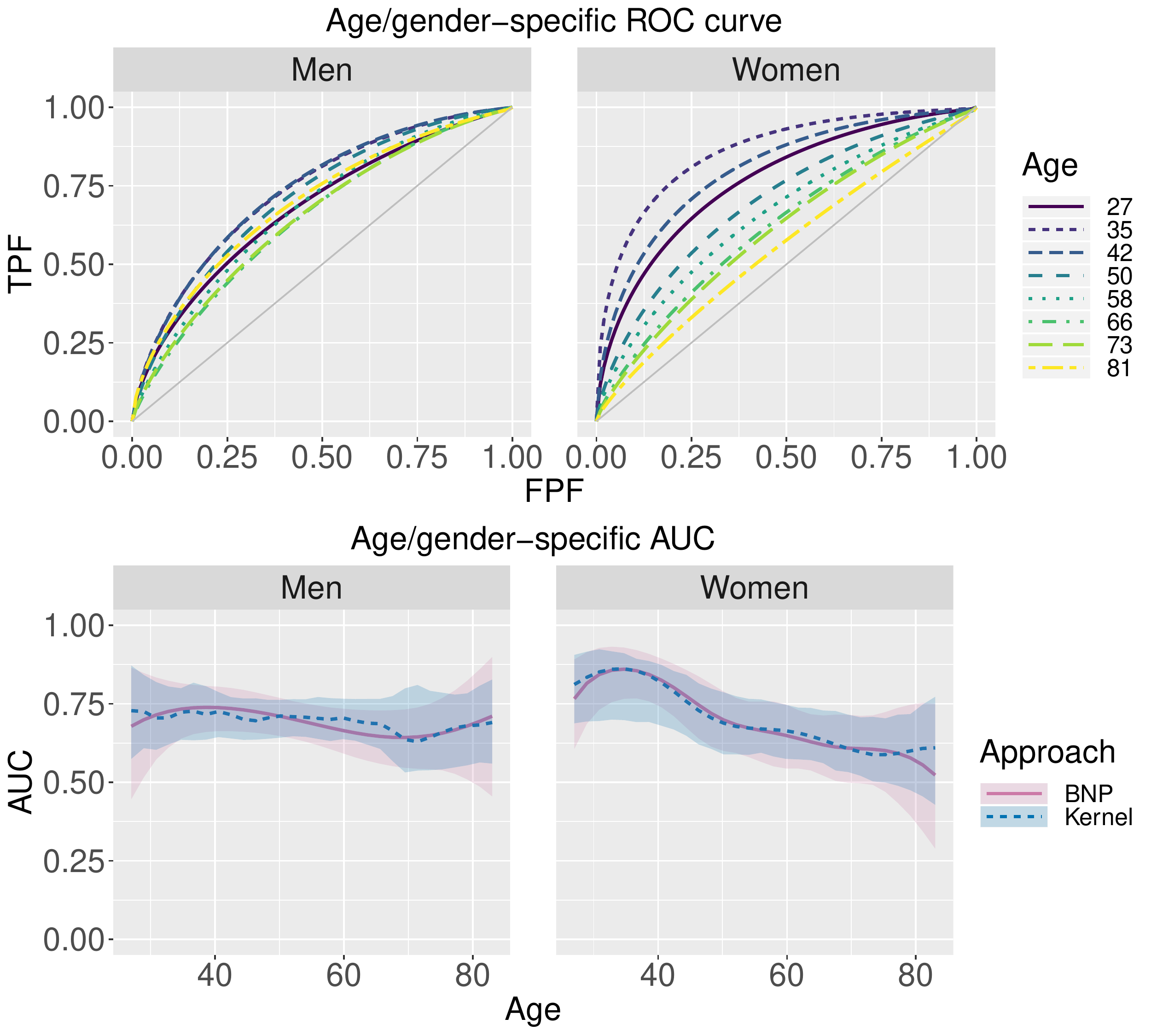}
\caption{Estimated age/gender-specific ROC curves using the approach of \cite{Inacio13} (top). Estimated age/gender-specific AUC. The continuous lines correspond to point estimates and the shaded regions correspond to the $95\%$ pointwise credible/confidence bands. Here BNP stands for the Bayesian nonparametric method of \cite{Inacio13} and Kernel for the approach of \cite{Rodriguez11}.}
\label{covariateresults}
\end{figure}

We revisit our example dataset and the aim now is to assess the effect of age and gender on the ability of HOMA-IR levels for predicting cardio metabolic-risk. In Figure \ref{covariateresults} (top) we present several ROC curves, obtained using the induced Bayesian nonparametric approach of \cite{Inacio13}, associated with different ages, for both men and women. While there is no substantial variation in the shape of the ROC curves in men, there is considerable differences for women (as already expectable given the conditional densities in Figure \ref{conditionaldensities}). To get deeper insight, in Figure \ref{covariateresults} (bottom) we depict the covariate-specific AUC for ages between  27 and 83 years old, which roughly correspond to the age interval where the two populations, for both men and women, had observations. Results are also shown for the kernel-based approach of \cite{Rodriguez11}, with the analysis in men and women conducted separately. It is also important to mention that for the approach of \cite{Inacio13} an interaction between age and gender was included. As foreseen, there is essentially no dynamic for the age-specific AUC in men. On the other hand, for women, the results suggest a decrease in the accuracy of HOMA-IR levels as age increases. Additionally, Figure 3 in the Supplementary Materials shows the age/gender-specific Youden index and associated age/gender-specific HOMA-IR optimal thresholds.

\section{ROC CURVES AND TIME (AND COVARIATES)}\label{timeROC}
Up to now we have been concerned about diagnosis. Yet, depending on the clinical circumstances, the aim and interest might involve prognosis rather than diagnosis. The main difference between diagnostic and prognostic settings is that the latter involves a time dimension. More specifically, in a prognostic setting the test outcome is measured at a given time (usually at baseline) and disease onset may occur at any time thereafter. As such, in prognosis, the true positive and negative fractions, and by consequence the ROC curve, are time dependent and may be calculated for different times. 

Here we only attempt to cover the main concepts, pointing the reader to the appropriate references about the estimation approaches.  With regard to notation, as before, let $Y$ be a continuous random variable denoting the test outcome and, additionally, let $T$, also a continuous random variable, denotes the time to disease onset. Further, let $D(t)$ be the disease status at time $t$, with $D(t) = 1$ indicating that disease onset is prior to time $t$, and $D(t) = 0$ otherwise. \cite{Heagerty05} proposed three definitions of the time-dependent true positive and negative fractions (which give rise to different definitions of the time-dependent ROC curve), namely, the \emph{cumulative} TPF and \emph{dynamic} TNF, the \emph{incident} TPF and \emph{dynamic} TNF, and the \emph{incident} TPF and \emph{static} TNF. These different definitions differ mainly in how disease and nondisease status are defined. We focus on the \emph{cumulative/dynamic} definition, where a diseased subject is any subject diagnosed between baseline (assumed to be the time $t=0$)  and time $t$ and a nondiseased subject is any individual free of disease at time $t$. From a practical point of view it has been argued \citep{Blanche13,Rodriguez16} that this is the most relevant definition, as clinicians often want to predict disease onset within a window of time rather than at a specific time (as in the \emph{incident} TPF) and the goal is also to distinguish nondiseased subjects at the end of such time window and not at a later pre-specified time (as implied by the \emph{static} FNF).

For a threshold $c$ and a given time $t$, \cite{Heagerty00} defined the cumulative true positive fraction $\text{TPF(c,t)}$ and the dynamic true negative fraction $\text{TNF}(c,t)$ by
\begin{align*}
\text{TPF}(c,t)&=\Pr(Y\geq c\mid D(t)=1)=\Pr(Y\geq c\mid T\leq t),\\
\text{TNF}(c,t)&=\Pr(Y< c\mid D(t)=0)=\Pr(Y< c\mid T >t).
\end{align*}
In words, the cumulative TPF is the probability that a subject has a test outcome equal or greater than $c$ among those individuals who developed the disease by time $t$, whereas the dynamic TNF is the probability that an individual has a test result less than $c$ among those who are disease free beyond that same time $t$. Under this definition, the sets of diseased and nondiseased subjects are changing over time and each individual might be in the nondiseased group at an earlier time and then in the diseased group at later times. The corresponding time-dependent ROC curve is defined as the plot of $\text{FPF}(c,t)$ versus $\text{TPF}(c,t)$ for all values of $c$, that is, $\{(\text{FPF}(c,t), \text{TPF}(c,t): c\in\mathbb{R}\}$. In analogy to Equation \ref{rocdef}, the time-dependent ROC curve can also be written as
\begin{equation*}
\text{ROC}(p, t)=\text{TPF}\{\text{FPF}^{-1}(p,t),t\},\qquad 0\leq p \leq 1,
\end{equation*}
where $\text{FPF}^{-1}(p,t)=\inf\{c\in\mathbb{R}: \text{FPF}(c,t)\leq p\}$. The AUC has been the preferred summary measure in the time-dependent context 
\begin{equation*}
\text{AUC}(t) = \int_{0}^{1}\text{ROC}(p, t)\text{d}p,
\end{equation*}
and it is worth noting that it also accepts a probabilistic intepretation
\begin{equation*}
\text{AUC}(t) = \Pr(Y_l>Y_m\mid D_l(t)=1, D_m(t)=0)=\Pr(Y_l>Y_m\mid T_l\leq t, T_m>t),
\end{equation*}
where $l$ and $m$ denote the indices of two randomly chosen subjects.

When it comes to estimating time-dependent ROC curves, one of the challenges is the (potential) presence of censoring. In practice, some subjects may be lost during the follow-up period, thus introducing right-censoring, and making it impossible to know if disease onset has happened before the time point $t$ for such subjects. Ignoring censoring might lead to biased estimates of the true positive and negative fractions. To address this issue, several approaches have been proposed to estimate the cumulative TPF, the dynamic TNF, and the corresponding ROC curve. The first proposal is due to \cite{Heagerty00} who developed estimators based on the Bayes' theorem and the Kaplan--Meier estimator of the survival function. The fact that this approach does not necessarily yield monotone true positive and negative fractions led the authors to propose an alternative approach based on a nearest neighbour estimator of the bivariate distribution of the test result and time to disease onset. Later \cite{Chambless06} proposed two alternative estimation methods for the TPF and TNF, one that deals with censoring by conditioning on the observed disease onset times as in the Kaplan-Meier estimator and another one that makes use of a Cox model. In turn, \cite{Uno07} and \cite{Hung10} both developed inverse probability of censoring weighting methods, while \cite{Martinez18} proposed an approach based on a bivariate kernel density estimator. From a Bayesian perspective, \cite{Zhao16} proposed a semiparametric approach that uses a single-weights dependent Dirichlet process mixture for modelling the conditional distribution of the time to disease onset given the test outcome. For recent overview articles on this topic see \cite{Blanche13} and \cite{Kamarudin17}, where the latter also surveys estimators proposed under the dynamic and incident definitions.
To conclude, we highlight that the inclusion of covariates, whenever available, in the time-dependent true positive and negative fractions, should also be done. The covariate-specific time-dependent TPF and TNF, for a covariate value $\mathbf{x}$, are given by
\begin{equation*}
\text{TPF}(c,t\mid\mathbf{x})=\Pr(Y\geq c\mid T\leq t, \mathbf{x}),\qquad \text{TNF}(c,t\mid\mathbf{x})=\Pr(Y< c\mid T> t, \mathbf{x}),
\end{equation*}
with the covariate-specific time-dependent ROC curve and AUC following in a similar fashion. The literature on the estimation of the covariate-specific time-dependent TPF and TNF is, by comparison, relatively scarce. Important references are \cite{Song08} and \cite{Rodriguez16}. 
\section{SOFTWARE}\label{software}
We start by making the disclaimer that we are not, by no means, doing an exhaustive review and that our focus is the \texttt{R} software. The package \texttt{pROC} (\url{https://CRAN.R-project.org/package=pROC}) provides a set of tools to visualise, smooth, and compare ROC curves, but covariate information cannot be explicitly taken into account. Packages \texttt{ROCRegression} and \texttt{npROCRegression} offer functions to estimate semiparametrically and nonparametrically,
under a frequentist framework and using both induced and direct methodologies, the covariate-specific ROC curve. In particular, \texttt{ROCRegression} (\url{https://bitbucket.org/mxrodriguez/rocregression}) implements the approaches of \cite{Faraggi03}, \cite{Pepe98}, \cite{Alonzo02}, and \cite{Cai04glm}, while \texttt{npROCRegression} (\url{https://CRAN.R-project.org/package=npROCRegression}) implements the approaches of \cite{Rodriguez11} and \cite{Rodriguez11new}. To the best of our knowledge, \texttt{ROCnReg} (\url{https://CRAN.R-project.org/package=ROCnReg}) is the only \texttt{R} package that allows conducting Bayesian inference for the ROC curve and related indices (including optimal thresholds) estimation. In particular, \texttt{ROCnReg} implements all four nonparametric approaches for ROC curve estimation described in Section \ref{rocest} and all induced approaches reviewed in Section \ref{covroc} for the estimation of the covariate-specific ROC curve. \texttt{ROCnReg} also offers routines for conducting inference about the covariate-adjusted ROC curve. All data analysis conducted in this article were obtained using \texttt{ROCnReg} \citep[for more details about the package see][]{MX20}. Also, the package \texttt{OptimalCutpoints} (\url{https://CRAN.R-project.org/package=OptimalCutpoints}) provides a collection of routines for point and interval estimation of optimal thresholds. Regarding estimation of the time-dependent ROC curve, the packages \texttt{survivalROC} (\url{https://CRAN.R-project.org/package=survivalROC}), \texttt{timeROC} (\url{https://CRAN.R-project.org/package=timeROC}), and \texttt{CondTimeROC} (\url{https://bitbucket.org/mxrodriguez/condtimeroc}) implement some of the approaches mentioned in Section \ref{timeROC}. 
\section{DISCUSSION AND FURTHER TOPICS}\label{discussion}
In this article we have reviewed from a high-level perspective some of the main aspects related to the statistical evaluation of medical tests. We have deliberately chosen to place special focus on the estimation of ROC curves, with and without covariates, with the case of time-dependent ROC curves being also covered. As a so broad area, many interesting topics have had necessarily to be left untouched and we briefly mention some of them below.

The available methodology for the study of the predictive values of continuous tests is far less extensive than the corresponding methodology for ROC curves. We mention the predictive receiver operating characteristic (pROC) curve proposed by \cite{Shiu08} for the joint assessment of the positive and negative predictive values. In an analogous way to the definition of the ROC curve, the authors defined the pROC as $\{1-\text{NPV}(c), \text{PPV}(c): c\in\mathbb{R}\}$. One possibility for its estimation is to make use of Equations \ref{ppv} and \ref{npv} (with the due adaptation that now in the continuous case all quantities are a function of the threshold) and then plug these estimates in the definition of the pROC curve, that is, in order to estimate the pROC curve we only need to estimate the corresponding TPF and FPFs and the prevalence of the disease. A covariate-specific ROC curve can be defined and estimated in a similar fashion. 

Although we have assumed that disease status is binary (disease versus nondisease), in clinical practice, physicians often face situations that require decisions among three (or even more) diagnostic alternatives. This is especially true for neurological disorders, where cognitive function usually declines from normal function to mild impairment, to severe impairment or dementia. ROC surfaces (and the volume under the surface and the generalised Youden index) have been proposed in the literature as an extension to the three-class case of ROC curve methodology \citep{Nakas04,Nakas10}. Parametric, semiparametric and nonparametric estimators do exist and we refer to \cite{Nakas14} for a recent overview. ROC surface regression, by opposed to the two-class counterpart, has received little attention, with \cite{Li12}, to the best of our knowledge, being the only contribution.

The existence of a gold standard test was assumed throughout this article, but this might not be the case for some diseases as, for instance, a definitive diagnosis of the Alzheimer's disease can only be made through autopsy after death. Approaches for estimating the ROC curve and the covariate-specific ROC curve in the absence of a gold standard test have been proposed, among others, by \cite{Branscum08} and \cite{Branscum15}.

Lastly, in our motivating example, the HOMA-IR levels, our diagnostic test/marker, was known and given. However, sometimes researchers do have access to multiple tests or biomarkers on individuals and interest in such cases might lie on how to best combine and transform this information onto a univariate score, to further use it to diagnose individuals. The topic of optimal combination of biomarkers using ROC analysis has received considerable attention in the literature (see, among many others, \citealt{Su93,Pepe06,Liu11}). Recently, methods that deal with optimal biomarker combination but with covariate adjustment have also been proposed \citep[e.g.][]{Liu13,Kim17}.

\section*{ACKNOWLEDGMENTS}
The work of V In\'acio was partially supported by FCT (Funda\c c\~ao para a Ci\^encia e a Tecnologia, Portugal), through the projects PTDC/MAT-STA/28649/2017 and \linebreak UIDB/00006/2020. MX Rodr\'iguez-\'Alvarez was funded by project MTM2017-82379-R (AEI/FEDER, UE), by the Basque Government through the BERC 2018-2021 program and by the Spanish Ministry of Science, Innovation, and Universities (BCAM Severo Ochoa accreditation SEV-2017-0718). 

\bibliographystyle{ar-style1}
\bibliography{references}

\begin{thebibliography}{}
\expandafter\ifx\csname natexlab\endcsname\relax\def\natexlab#1{#1}\fi

\bibitem[Alonzo \& Pepe(2002)]{Alonzo02}
Alonzo TA, Pepe MS. 2002.
Distibution‐-free {ROC} analysis using binary regression techniques.
\textit{Biostatistics} 3:421--432

\bibitem[Bamber(1975)]{Bamber1975}
Bamber D. 1975.
The area above the ordinal dominance graph and the area below the receiver
  operating characteristic graph.
\textit{Journal of Mathematical Psychology} 12:387--415

\bibitem[Blanche et~al.(2013)Blanche, Dartigues \& Jacqmin-Gadda]{Blanche13}
Blanche P, Dartigues JF, Jacqmin-Gadda H. 2013.
Review and comparison of {ROC} curve estimators for a time-dependent outcome
  with marker-dependent censoring.
\textit{Biometrical Journal} 55:687--704

\bibitem[Branscum et~al.(2015)Branscum, Johnson, Hanson \& Baron]{Branscum15}
Branscum AJ, Johnson WO, Hanson TE, Baron AT. 2015.
Flexible regression models for {ROC} and risk analysis, with or without a gold
  standard.
\textit{Statistics in Medicine} 34:3997--4015

\bibitem[Branscum et~al.(2008)Branscum, Johnson, Hanson \& Gardner]{Branscum08}
Branscum AJ, Johnson WO, Hanson TE, Gardner IA. 2008.
Bayesian semiparametric {ROC} curve estimation and disease diagnosis.
\textit{Statistics in Medicine} 27:2474--2496

\bibitem[Broemeling(2016)]{Broemeling16}
Broemeling LD. 2016.
\emph{Advanced {B}ayesian {M}ethods for {M}edical {T}est {A}ccuracy}.
Chapman \& Hall/CRC Press

\bibitem[Brownie et~al.(1986)Brownie, Habicht \& Cogill]{Brownie86}
Brownie C, Habicht JP, Cogill B. 1986.
Comparing indicators of health or nutritional status.
\textit{American Journal of Epidemiology} 124:1031--1044

\bibitem[Cai(2004)]{Cai04glm}
Cai T. 2004.
Semi-parametric {ROC} regression analysis with placement values.
\textit{Biostatistics} 5:45--60

\bibitem[Cai \& Dodd(2008)]{Cai08}
Cai T, Dodd LE. 2008.
Regression analysis for the partial area under the {ROC} curve.
\textit{Statistica Sinica} 18:817--836

\bibitem[Cai \& Moskowitz(2004)]{Cai04}
Cai T, Moskowitz CS. 2004.
Semi-parametric estimation of the binormal {ROC} curve for a continuous
  diagnostic test.
\textit{Biostatistics} 5:573--586

\bibitem[Cai \& Pepe(2002)]{Cai02}
Cai T, Pepe MS. 2002.
Semiparametric receiver operating characteristic analysis to evaluate
  biomarkers for disease.
\textit{Journal of the American Statistical Association} 97:1099--1107

\bibitem[Chambless \& Diao(2006)]{Chambless06}
Chambless LE, Diao G. 2006.
Estimation of time-dependent area under the {ROC} curve for long-term risk
  prediction.
\textit{Statistics in Medicine} 25:3474--3486

\bibitem[Dodd \& Pepe(2003{\natexlab{a}})]{Dodd03b}
Dodd LE, Pepe MS. 2003{\natexlab{a}}.
Partial {AUC} estimation and regression.
\textit{Biometrics} 59:614--623

\bibitem[Dodd \& Pepe(2003{\natexlab{b}})]{Dodd03a}
Dodd LE, Pepe MS. 2003{\natexlab{b}}.
Semiparametric regression for the area under the receiver operating
  characteristic curve.
\textit{Journal of the American Statistical Association} 98:409--417

\bibitem[Dorfman \& Alf(1969)]{Dorfman69}
Dorfman DD, Alf E. 1969.
Maximum-likelihood estimation of parameters of signal-detection theory and
  determination of confidence intervals—rating-method data.
\textit{Journal of Mathematical Psychology} 6:487--496

\bibitem[Erkanli et~al.(2006)Erkanli, Sung, Jane~Costello \& Angold]{Erkanli06}
Erkanli A, Sung M, Jane~Costello E, Angold A. 2006.
Bayesian semi-parametric {ROC} analysis.
\textit{Statistics in Medicine} 25:3905--3928

\bibitem[Escobar \& West(1995)]{Escobar95}
Escobar MD, West M. 1995.
Bayesian density estimation and inference using mixtures.
\textit{Journal of the American Statistical Association} 90:577--588

\bibitem[Fan \& Gijbels(1996)]{Fan96}
Fan J, Gijbels I. 1996.
\emph{{L}ocal {P}olynomial {M}odelling and {I}ts {A}pplications}.
Chapman \& Hall/CRC Press

\bibitem[Faraggi(2003)]{Faraggi03}
Faraggi D. 2003.
Adjusting receiver operating characteristic curves and related indices for
  covariates.
\textit{Journal of the Royal Statistical Society: Series D} 52:179--192

\bibitem[Fawcett(2006)]{Fawcett2006}
Fawcett T. 2006.
An introduction to {ROC} analysis.
\textit{Pattern Recognition Letters} 27:861--874

\bibitem[Ferguson(1973)]{Ferguson73}
Ferguson TS. 1973.
A {B}ayesian analysis of some nonparametric problems.
\textit{The Annals of Statistics} 1:209--230

\bibitem[Fluss et~al.(2005)Fluss, Faraggi \& Reiser]{Fluss05}
Fluss R, Faraggi D, Reiser B. 2005.
Estimation of the {Y}ouden index and its associated cutoff point.
\textit{Biometrical Journal} 47:458--472

\bibitem[Gayoso-Diz et~al.(2013)Gayoso-Diz, Otero-Gonz{\'a}lez,
  Rodriguez-Alvarez, Gude, Garc{\'\i}a et~al.]{Gayoso13}
Gayoso-Diz P, Otero-Gonz{\'a}lez A, Rodriguez-Alvarez MX, Gude F, Garc{\'\i}a
  F, et~al. 2013.
Insulin resistance ({HOMA-IR}) cut-off values and the metabolic syndrome in a
  general adult population: effect of gender and age: {EPIRCE} cross-sectional
  study.
\textit{BMC Endocrine Disorders} 13:47

\bibitem[Gneiting \& Vogel(2018)]{Gneiting18}
Gneiting T, Vogel P. 2018.
Receiver {O}perating {C}haracteristic ({ROC}) {C}urves.
\textit{arXiv:1809.04808}

\bibitem[Goddard \& Hinberg(1990)]{Goddard90}
Goddard M, Hinberg I. 1990.
Receiver operator characteristic ({ROC}) curves and non-normal data: an
  empirical study.
\textit{Statistics in Medicine} 9:325--337

\bibitem[Gon{\c{c}}alves et~al.(2014)Gon{\c{c}}alves, Subtil, Oliveira \&
  Bermudez]{Goncalves2014}
Gon{\c{c}}alves L, Subtil A, Oliveira MR, Bermudez P. 2014.
{ROC} curve estimation: {A}n overview.
\textit{REVSTAT--Statistical Journal} 12:1--20

\bibitem[Gonz{\'a}lez-Manteiga et~al.(2011)Gonz{\'a}lez-Manteiga,
  Pardo-Fern{\'a}ndez \& Van~Keilegom]{Gonzalez11}
Gonz{\'a}lez-Manteiga W, Pardo-Fern{\'a}ndez JC, Van~Keilegom I. 2011.
{ROC} curves in non-parametric location-scale regression models.
\textit{Scandinavian Journal of Statistics} 38:169--184

\bibitem[Gu \& Ghosal(2009)]{Gu09}
Gu J, Ghosal S. 2009.
Bayesian {ROC} curve estimation under binormality using a rank likelihood.
\textit{Journal of Statistical Planning and Inference} 139:2076--2083

\bibitem[Gu et~al.(2008)Gu, Ghosal \& Roy]{Gu08}
Gu J, Ghosal S, Roy A. 2008.
Bayesian bootstrap estimation of {ROC} curve.
\textit{Statistics in Medicine} 27:5407--5420

\bibitem[Hanley(1996)]{Hanley96}
Hanley JA. 1996.
The use of the ‘binormal’model for parametric {ROC} analysis of
  quantitative diagnostic tests.
\textit{Statistics in Medicine} 15:1575--1585

\bibitem[Heagerty et~al.(2000)Heagerty, Lumley \& Pepe]{Heagerty00}
Heagerty PJ, Lumley T, Pepe MS. 2000.
Time-dependent {ROC} curves for censored survival data and a diagnostic marker.
\textit{Biometrics} 56:337--344

\bibitem[Heagerty \& Zheng(2005)]{Heagerty05}
Heagerty PJ, Zheng Y. 2005.
Survival model predictive accuracy and {ROC} curves.
\textit{Biometrics} 61:92--105

\bibitem[Hsieh \& Turnbull(1996)]{Hsieh96}
Hsieh F, Turnbull BW. 1996.
Nonparametric and semiparametric estimation of the receiver operating
  characteristic curve.
\textit{The Annals of Statistics} 24:25--40

\bibitem[Hung \& Chiang(2010)]{Hung10}
Hung H, Chiang CT. 2010.
Optimal composite markers for time-dependent receiver operating characteristic
  curves with censored survival data.
\textit{Scandinavian Journal of Statistics} 37:664--679

\bibitem[In{\'a}cio~de Carvalho et~al.(2013)In{\'a}cio~de Carvalho, Jara,
  Hanson \& de~Carvalho]{Inacio13}
In{\'a}cio~de Carvalho V, Jara A, Hanson TE, de~Carvalho M. 2013.
Bayesian nonparametric {ROC} regression modeling.
\textit{Bayesian Analysis} 8:623--646

\bibitem[In{\'a}cio~de Carvalho \& Rodr{\'i}guez-{\'A}lvarez(2018)]{Inacio18}
In{\'a}cio~de Carvalho V, Rodr{\'i}guez-{\'A}lvarez MX. 2018.
Bayesian nonparametric inference for the covariate-adjusted {ROC} curve.
\textit{arXiv:1806.00473}

\bibitem[{International Diabetes Federation}(2006)]{IDF20}
{International Diabetes Federation}. 2006.
{IDF} {C}onsensus {W}orldwide {D}efinition of the {M}etabolic {S}yndrome.
\url{https://www.idf.org/e-library/consensus-statements/60-idfconsensus-worldwide-definitionof-the-metabolic-syndrome.html}
  (accessed March 30, 2020)

\bibitem[Ishwaran \& James(2002)]{Ishwaran02}
Ishwaran H, James LF. 2002.
Approximate {D}irichlet process computing in finite normal mixtures: smoothing
  and prior information.
\textit{Journal of Computational and Graphical Statistics} 11:508--532

\bibitem[Janes \& Pepe(2008{\natexlab{a}})]{Janes08a}
Janes H, Pepe MS. 2008{\natexlab{a}}.
Adjusting for covariates in studies of diagnostic, screening, or prognostic
  markers: An old concept in a new setting.
\textit{American Journal of Epidemiology} 168:89--97

\bibitem[Janes \& Pepe(2008{\natexlab{b}})]{Janes08b}
Janes H, Pepe MS. 2008{\natexlab{b}}.
Matching in studies of classification accuracy: implications for analysis,
  efficiency, and assessment of incremental value.
\textit{Biometrics} 64:1--9

\bibitem[Janes \& Pepe(2009)]{Janes09}
Janes H, Pepe MS. 2009.
Adjusting for covariate effects on classification accuracy using the
  covariate-adjusted receiver operating characteristic curve.
\textit{Biometrika} 96:371--382

\bibitem[Kamarudin et~al.(2017)Kamarudin, Cox \& Kolamunnage-Dona]{Kamarudin17}
Kamarudin AN, Cox T, Kolamunnage-Dona R. 2017.
Time-dependent {ROC} curve analysis in medical research: current methods and
  applications.
\textit{BMC Medical Research Methodology} 17:53

\bibitem[Kim \& Huang(2017)]{Kim17}
Kim S, Huang Y. 2017.
Combining biomarkers for classification with covariate adjustment.
\textit{Statistics in Medicine} 36:2347--2362

\bibitem[Krzanowski \& Hand(2009)]{Krzanowski09}
Krzanowski WJ, Hand DJ. 2009.
\emph{{ROC} {C}urves for {C}ontinuous {D}ata}.
Chapman \& Hall/CRC Press

\bibitem[Li et~al.(2012)Li, Zhou \& Fine]{Li12}
Li J, Zhou X, Fine JP. 2012.
A regression approach to {ROC} surface, with applications to {A}lzheimer’s
  disease.
\textit{Science China Mathematics} 55:1583--1595

\bibitem[Lin et~al.(2012)Lin, Zhou \& Li]{Lin12}
Lin H, Zhou XH, Li G. 2012.
A direct semiparametric receiver operating characteristic curve regression with
  unknown link and baseline functions.
\textit{Statistica Sinica} 22:1427--1456

\bibitem[Liu et~al.(2011)Liu, Liu \& Halabi]{Liu11}
Liu C, Liu A, Halabi S. 2011.
A min--max combination of biomarkers to improve diagnostic accuracy.
\textit{Statistics in Medicine} 30:2005--2014

\bibitem[Liu \& Zhou(2013)]{Liu13}
Liu D, Zhou XH. 2013.
{ROC} analysis in biomarker combination with covariate adjustment.
\textit{Academic Radiology} 20:874--882

\bibitem[Lloyd(1998)]{Lloyd98}
Lloyd CJ. 1998.
Using smoothed receiver operating characteristic curves to summarize and
  compare diagnostic systems.
\textit{Journal of the American Statistical Association} 93:1356--1364

\bibitem[L{\'o}pez-de Ullibarri et~al.(2008)L{\'o}pez-de Ullibarri, Cao,
  Cadarso-Su{\'a}rez \& Lado]{Lopez08}
L{\'o}pez-de Ullibarri I, Cao R, Cadarso-Su{\'a}rez C, Lado MJ. 2008.
Nonparametric estimation of conditional {ROC} curves: {A}pplication to
  discrimination tasks in computerized detection of early breast cancer.
\textit{Computational Statistics \& Data Analysis} 52:2623--2631

\bibitem[Mart{\'\i}nez-Camblor et~al.(2017)Mart{\'\i}nez-Camblor, Corral, Rey,
  Pascual \& Cernuda-Moroll{\'o}n]{Martinez2017}
Mart{\'\i}nez-Camblor P, Corral N, Rey C, Pascual J, Cernuda-Moroll{\'o}n E.
  2017.
Receiver operating characteristic curve generalization for non-monotone
  relationships.
\textit{Statistical Methods in Medical Research} 26:113--123

\bibitem[Mart{\'\i}nez-Camblor \& Pardo-Fern{\'a}ndez(2018)]{Martinez18}
Mart{\'\i}nez-Camblor P, Pardo-Fern{\'a}ndez JC. 2018.
Smooth time-dependent receiver operating characteristic curve estimators.
\textit{Statistical Methods in Medical Research} 27:651--674

\bibitem[Metz(1978)]{Metz78}
Metz CE. 1978.
Basic principles of {ROC} analysis, In \textit{Seminars in Nuclear Medicine},
  vol.~8,\ pp.  283--298, WB Saunders

\bibitem[Metz(1986)]{Metz86}
Metz CE. 1986.
{ROC} methodology in radiologic imaging.
\textit{Investigative Radiology} 21:720--733

\bibitem[Metz et~al.(1998)Metz, Herman \& Shen]{Metz98}
Metz CE, Herman BA, Shen JH. 1998.
Maximum likelihood estimation of receiver operating characteristic ({ROC})
  curves from continuously-distributed data.
\textit{Statistics in Medicine} 17:1033--1053

\bibitem[Nakas(2014)]{Nakas14}
Nakas CT. 2014.
Developments in {ROC} surface analysis and assessment of diagnostic markers in
  three-class classification problems.
\textit{REVSTAT--Stat J} 12:43--65

\bibitem[Nakas et~al.(2010)Nakas, Alonzo \& Yiannoutsos]{Nakas10}
Nakas CT, Alonzo TA, Yiannoutsos CT. 2010.
Accuracy and cut-off point selection in three-class classification problems
  using a generalization of the youden index.
\textit{Statistics in Medicine} 29:2946--2955

\bibitem[Nakas \& Yiannoutsos(2004)]{Nakas04}
Nakas CT, Yiannoutsos CT. 2004.
Ordered multiple-class {ROC} analysis with continuous measurements.
\textit{Statistics in Medicine} 23:3437--3449

\bibitem[Otero et~al.(2010)Otero, de~Francisco, Gayoso \& Garc\'ia]{Otero10}
Otero A, de~Francisco A, Gayoso P, Garc\'ia F. 2010.
Prevalence of chronic renal disease in {S}pain: results of the {EPIRCE} study.
\textit{Nefrolog\'ia : publicaci\'on oficial de la Sociedad Espa\~nola
  Nefrologia} 30:78--86

\bibitem[Otero et~al.(2005)Otero, Gayoso, Garcia \& de~Francisco]{Otero05}
Otero A, Gayoso P, Garcia F, de~Francisco AL. 2005.
Epidemiology of chronic renal disease in the {G}alician population: results of
  the pilot {S}panish {EPIRCE} study.
\textit{Kidney international. Supplement} :S16--9

\bibitem[Pardo~Fern{\'a}ndez et~al.(2014)Pardo~Fern{\'a}ndez,
  Rodr{\'\i}guez~{\'A}lvarez \& Van~Keilegom]{Pardo2014}
Pardo~Fern{\'a}ndez JC, Rodr{\'\i}guez~{\'A}lvarez MX, Van~Keilegom I. 2014.
A review on {ROC} curves in the presence of covariates 12:21--41

\bibitem[Pepe(1998)]{Pepe98}
Pepe MS. 1998.
Three approaches to regression analysis of receiver operating characteristic
  curves for continuous test results.
\textit{Biometrics} 54:124--135

\bibitem[Pepe(2000)]{Pepe00}
Pepe MS. 2000.
An interpretation for the {ROC} curve and inference using {GLM} procedures.
\textit{Biometrics} 56:352--359

\bibitem[Pepe(2003)]{Pepe03}
Pepe MS. 2003.
\emph{{T}he {S}tatistical {E}valuation of {M}edical {T}ests for
  {C}lassification and {P}rediction}.
Oxford University Press

\bibitem[Pepe \& Cai(2004)]{Pepe04}
Pepe MS, Cai T. 2004.
The analysis of placement values for evaluating discriminatory measures.
\textit{Biometrics} 60:528--535

\bibitem[Pepe et~al.(2006)Pepe, Cai \& Longton]{Pepe06}
Pepe MS, Cai T, Longton G. 2006.
Combining predictors for classification using the area under the receiver
  operating characteristic curve.
\textit{Biometrics} 62:221--229

\bibitem[{\texttt{R} Core Team}(2020)]{R20}
{\texttt{R} Core Team}. 2020.
\texttt{R}: {A} language and environment for statistical computing.
\texttt{R} Foundation for Statistical Computing, Vienna, Austria

\bibitem[Rodr{\'\i}guez \& Mart{\'\i}nez(2014)]{Rodriguez14}
Rodr{\'\i}guez A, Mart{\'\i}nez JC. 2014.
Bayesian semiparametric estimation of covariate-dependent {ROC} curves.
\textit{Biostatistics} 15:353--369

\bibitem[Rodriguez-Alvarez \& In\'acio(2020)]{MX20}
Rodriguez-Alvarez MX, In\'acio V. 2020.
{ROCnReg}: An {R} package for receiver operating characteristic curve inference
  with and without covariate information.
\textit{arXiv:2003.13111}

\bibitem[Rodr{\'\i}guez-{\'A}lvarez et~al.(2016)Rodr{\'\i}guez-{\'A}lvarez,
  Meira-Machado, Abu-Assi \& Raposeiras-Roub{\'\i}n]{Rodriguez16}
Rodr{\'\i}guez-{\'A}lvarez MX, Meira-Machado L, Abu-Assi E,
  Raposeiras-Roub{\'\i}n S. 2016.
Nonparametric estimation of time-dependent {ROC} curves conditional on a
  continuous covariate.
\textit{Statistics in Medicine} 35:1090--1102

\bibitem[Rodr{\'\i}guez-{\'A}lvarez
  et~al.(2011{\natexlab{a}})Rodr{\'\i}guez-{\'A}lvarez, Roca-Pardi{\~n}as \&
  Cadarso-Su{\'a}rez]{Rodriguez11new}
Rodr{\'\i}guez-{\'A}lvarez MX, Roca-Pardi{\~n}as J, Cadarso-Su{\'a}rez C.
  2011{\natexlab{a}}.
A new flexible direct {ROC} regression model: Application to the detection of
  cardiovascular risk factors by anthropometric measures.
\textit{Computational Statistics \& Data Analysis} 55:3257--3270

\bibitem[Rodr{\'\i}guez-{\'A}lvarez
  et~al.(2011{\natexlab{b}})Rodr{\'\i}guez-{\'A}lvarez, Roca-Pardi{\~n}as \&
  Cadarso-Su{\'a}rez]{Rodriguez11}
Rodr{\'\i}guez-{\'A}lvarez MX, Roca-Pardi{\~n}as J, Cadarso-Su{\'a}rez C.
  2011{\natexlab{b}}.
{ROC} curve and covariates: extending induced methodology to the non-parametric
  framework.
\textit{Statistics and Computing} 21:483--499

\bibitem[Sethuraman(1994)]{Sethuraman94}
Sethuraman J. 1994.
A constructive definition of {D}irichlet priors.
\textit{Statistica Sinica} 4:639--650

\bibitem[Shiu \& Gatsonis(2008)]{Shiu08}
Shiu SY, Gatsonis C. 2008.
The predictive receiver operating characteristic curve for the joint assessment
  of the positive and negative predictive values.
\textit{Philosophical Transactions of the Royal Society A} 366:2313--2333

\bibitem[Silverman(1986)]{Silverman86}
Silverman BW. 1986.
\emph{{D}ensity {E}stimation for {S}tatistics and {D}ata {A}nalysis}.
Chapman \&Hall/CRC Press

\bibitem[Song \& Zhou(2008)]{Song08}
Song X, Zhou XH. 2008.
A semiparametric approach for the covariate specific {ROC} curve with survival
  outcome.
\textit{Statistica Sinica} 18:947--965

\bibitem[Su \& Liu(1993)]{Su93}
Su JQ, Liu JS. 1993.
Linear combinations of multiple diagnostic markers.
\textit{Journal of the American Statistical Association} 88:1350--1355

\bibitem[Swets(1986)]{Swets86}
Swets JA. 1986.
Indices of discrimination or diagnostic accuracy: their {ROC}s and implied
  models.
\textit{Psychological Bulletin} 99:100--117

\bibitem[Uno et~al.(2007)Uno, Cai, Tian \& Wei]{Uno07}
Uno H, Cai T, Tian L, Wei LJ. 2007.
Evaluating prediction rules for t-year survivors with censored regression
  models.
\textit{Journal of the American Statistical Association} 102:527--537

\bibitem[Xu et~al.(2014)Xu, Wang \& Fang]{Xu14}
Xu T, Wang J, Fang Y. 2014.
A model-free estimation for the covariate-adjusted {Y}ouden index and its
  associated cut-point.
\textit{Statistics in Medicine} 33:4963--4974.

\bibitem[Yao et~al.(2010)Yao, Craiu \& Reiser]{Yao10}
Yao F, Craiu RV, Reiser B. 2010.
Nonparametric covariate adjustment for receiver operating characteristic
  curves.
\textit{Canadian Journal of Statistics} 38:27--46

\bibitem[Youden(1950)]{Youden50}
Youden WJ. 1950.
Index for rating diagnostic tests.
\textit{Cancer} 3:32--35

\bibitem[Zhao et~al.(2016)Zhao, Feng, Chen \& Taylor]{Zhao16}
Zhao L, Feng D, Chen G, Taylor JM. 2016.
A unified {B}ayesian semiparametric approach to assess discrimination ability
  in survival analysis.
\textit{Biometrics} 72:554--562

\bibitem[Zheng \& Heagerty(2004)]{Zheng04}
Zheng Y, Heagerty PJ. 2004.
Semiparametric estimation of time-dependent {ROC} curves for longitudinal
  marker data.
\textit{Biostatistics} 5:615--632

\bibitem[Zhou \& Harezlak(2002)]{Zhou02}
Zhou XH, Harezlak J. 2002.
Comparison of bandwidth selection methods for kernel smoothing of {ROC} curves.
\textit{Statistics in Medicine} 21:2045--2055

\bibitem[Zhou et~al.(2011)Zhou, McClish \& Obuchowski]{Zhou11}
Zhou XH, McClish DK, Obuchowski NA. 2011.
\emph{Statistical {M}ethods in {D}iagnostic {M}edicine}.
John Wiley \& Sons

\bibitem[Zou \& Hall(2000)]{Zou00}
Zou KH, Hall W. 2000.
Two transformation models for estimating an {ROC} curve derived from continuous
  data.
\textit{Journal of Applied Statistics} 27:621--631

\bibitem[Zou et~al.(1997)Zou, Hall \& Shapiro]{Zou97}
Zou KH, Hall W, Shapiro DE. 1997.
Smooth non-parametric receiver operating characteristic ({ROC}) curves for
  continuous diagnostic tests.
\textit{Statistics in Medicine} 16:2143--2156

\bibitem[Zou et~al.(1998)Zou, Tempany, Fielding \& Silverman]{Zou98}
Zou KH, Tempany CM, Fielding JR, Silverman SG. 1998.
Original smooth receiver operating characteristic curve estimation from
  continuous data: statistical methods for analyzing the predictive value of
  spiral {CT} of ureteral stones.
\textit{Academic Radiology} 5:680--687

\end{thebibliography}

\newpage

\section*{SUPPLEMENTARY MATERIALS}
\setcounter{figure}{0}  
\setcounter{table}{0}  
Here we provide supplementary figures and tables to the main document.

\begin{figure}[h]
\begin{center}
\includegraphics[width=13.5cm]{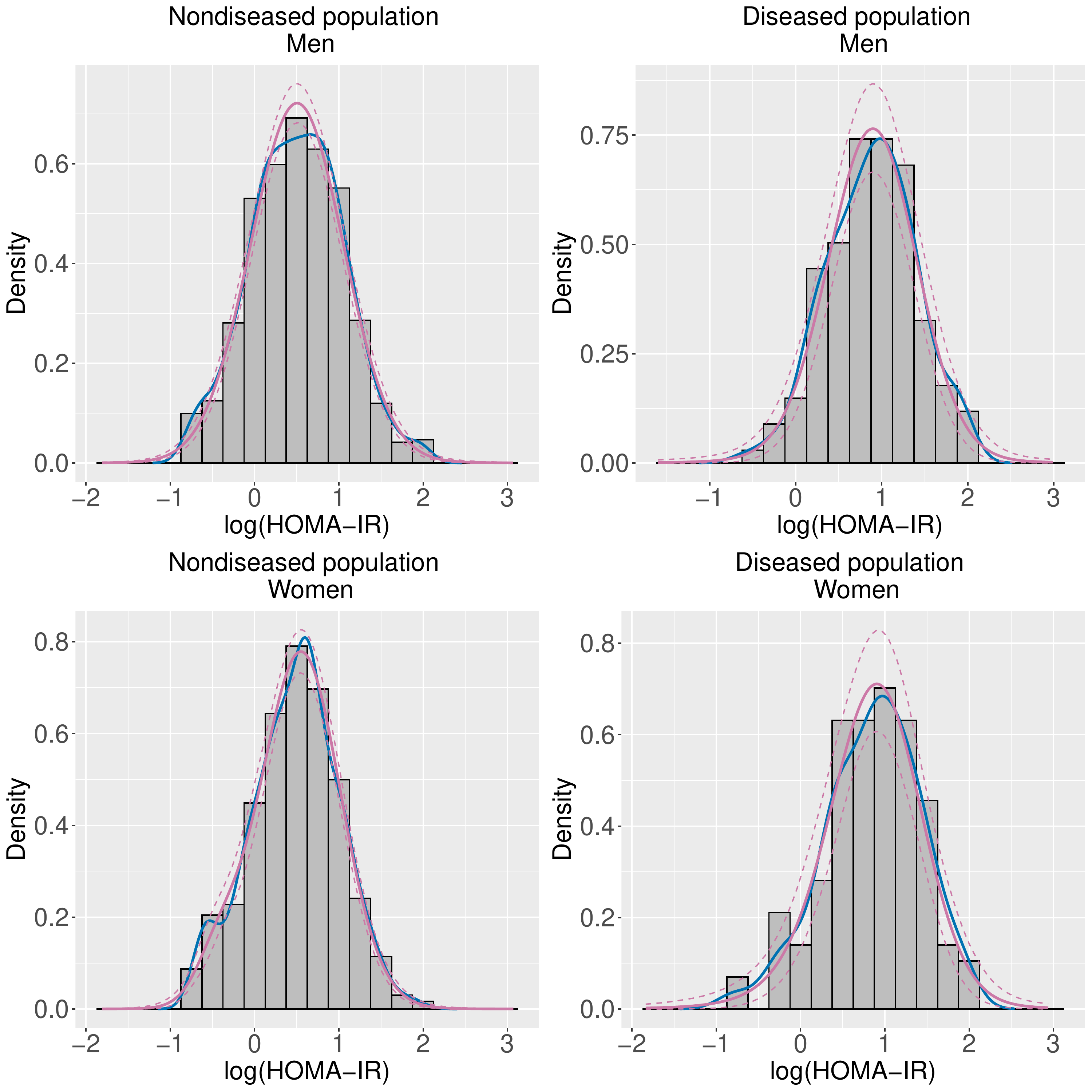}
\caption{Histograms of the $\log$ HOMA IR levels along with the estimated densities produced by a Dirichlet process mixture of normals model (solid pink line, with the dashed pink lines representing the pointwise 95\% credible band) and by a kernel method (normal kernel and bandwidth selected by Silverman's rule of thumb) (solid blue line).}
\end{center}
\end{figure}

\begin{figure}[h]
\begin{center}
\includegraphics[width=16cm]{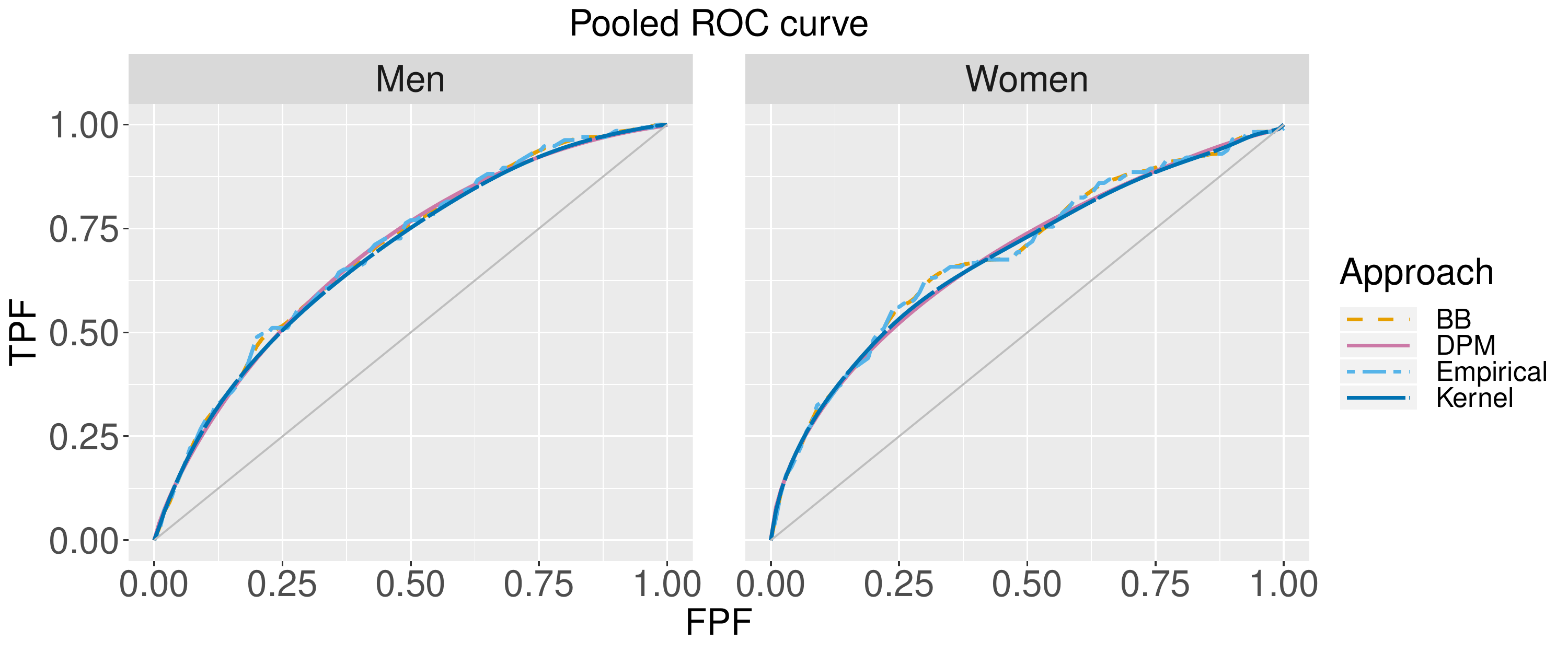}
\caption{Estimated ROC curve. Here BB stands for the Bayesian bootstrap method (Gu et al. 2008) and DPM for the Dirichlet process mixture of normals model (Erkanli et al. 2006).}
\end{center}
\end{figure}

\begin{table}[h]
\begin{center}
\begin{tabular}{ccccc}
\hline
& \multicolumn{2}{c}{Women} & \multicolumn{2}{c}{Men} \\\hline           
& \begin{footnotesize}Youden index\end{footnotesize} &  \begin{footnotesize}$\log$ HOMA-IR optimal threshold \end{footnotesize} & \begin{footnotesize} Youden index\end{footnotesize} &\begin{footnotesize}  $\log$ HOMA-IR optimal threshold\end{footnotesize}\\
Empirical &$0.325$& $0.742$    & $0.292$ &$0.718$      \\
Kernel &$0.283$& $0.804$    & $0.265$ & $0.728$      \\
DPM &$0.277$ $(0.197,0.356)$& $0.779$ $(0.660,0.894)$    & $0.282$ $(0.210,0.353)$ &$0.666$ $(0.557,0.779)$      \\
BB &$0.338$ $(0.249,0.427)$& $0.781$ $(0.695,0.913)$    & $0.315$ $(0.237,0.390)$ & $0.757$ $(0.400,0.962)$      \\
\hline
\end{tabular}
\end{center}
\caption{Estimated Youden index and associated log Homa-IR optimal threshold. For the Bayesian approaches, the numbers in brackets are the $95\%$ credible intervals. Here BB stands for the Bayesian bootstrap method (Gu et al. 2008) and DPM for the Dirichlet process mixture of normals model (Erkanli et al. 2006).}
\end{table}

\begin{figure}[h]
\begin{center}
\includegraphics[width=13.5cm]{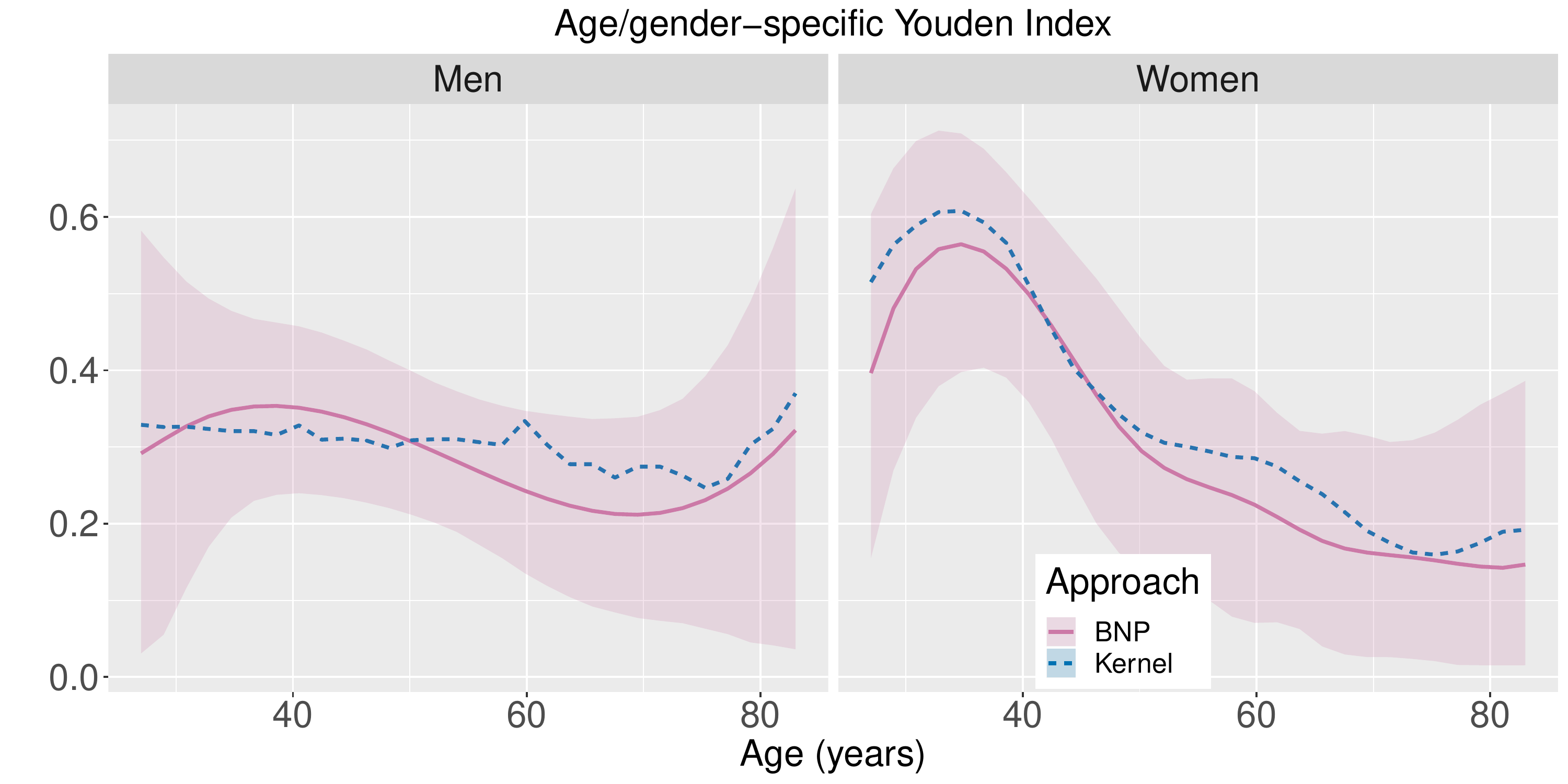}
\includegraphics[width=13.5cm]{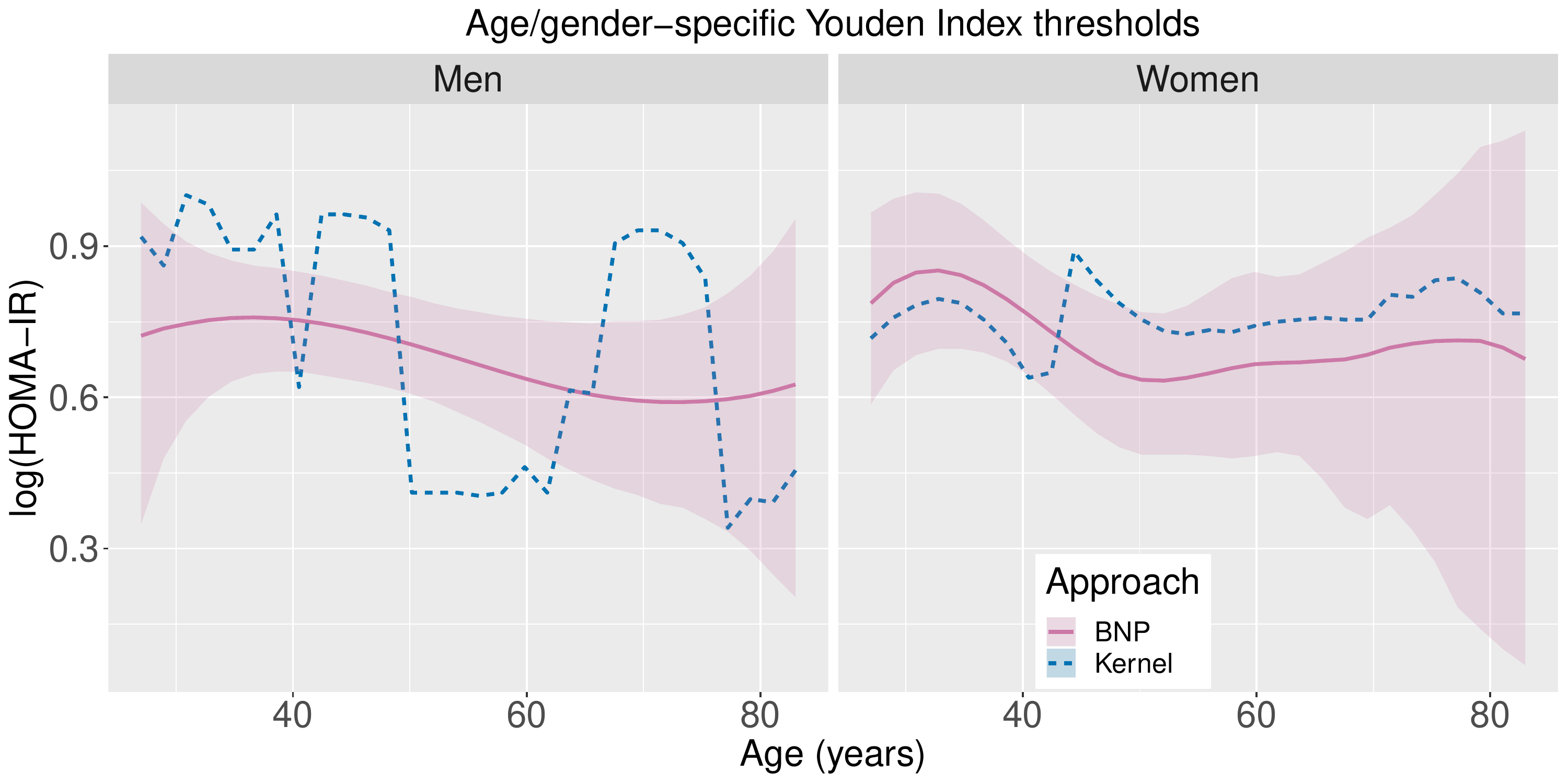}\\
\caption{Estimated age/gender-specific Youden index and associated $\log$ HOMA-IR optimal thresholds. The continuous lines correspond to point estimates and the shaded region correspond to the $95\%$ pointwise credible band. Here BNP stands for the Bayesian nonparametric method of In\'acio de Carvalho et al. (2013) and Kernel for the approach of Rodr\'iguez-\'Alvarez et al. (2011). }
\end{center}
\end{figure}

\end{document}